%% file: main.tex
\def\year{2019}\relax
\documentclass[letterpaper,preprint]{article} 
\usepackage{aaai19}  
\usepackage{times}  
\usepackage{helvet} 
\usepackage{courier}  
\usepackage[hyphens]{url}  
\usepackage{graphicx} 
\def\UrlFont{\rm}  
\usepackage{graphicx}  
\nocopyright
\title{Improved Forecasting of Cryptocurrency Price using Social Signals}
\author{Maria Glenski\\
Data Science and Analytics\\
Pacific Northwest National Laboratory\\
maria.glenski@pnnl.gov
\And
 Tim Weninger\\
 Computer Science and Engineering\\
University of Notre Dame\\
tweninge@nd.edu
\And
 Svitlana Volkova\\
Data Science and Analytics\\
Pacific Northwest National Laboratory\\
svitlana.volkova@pnnl.gov
}

\usepackage{amsmath}

\usepackage{times}
\usepackage{latexsym}

\usepackage{url}

\usepackage{xcolor}

\usepackage{tikz} 
\usepackage{pgfplots}
\usepgfplotslibrary{dateplot}
\pgfplotsset{compat=1.14}

\usepackage{algorithm}
\usepackage[noend]{algpseudocode}

\usepackage{cleveref}[2012/02/15]

\crefformat{footnote}{#2\footnotemark[#1]#3}
\usepackage{pgfplotstable}
\usepgfplotslibrary{fillbetween}

\pgfplotsset{
	discard if not/.style 2 args={
		x filter/.code={
			\edef\tempa{\thisrow{#1}}
			\edef\tempb{#2}
			\ifx\tempa\tempb
			\else
			
			\fi
		}
	}
}

\usepackage{booktabs} 
\usepackage{multirow}

\newcommand*{\ie}{{\em i.e.,~}}
\newcommand*{\eg}{{\em e.g.,~}}
\newcommand*{\etal}{et al.~}
\newcommand*{\ignore}[1]{} 
\newcommand*{\nop}[1]{}

\usepackage{xcolor}

\colorlet{cluster0}{blue}
\colorlet{cluster1}{blue!80!white}
\colorlet{cluster2}{blue!60!white}
\colorlet{cluster3}{blue!50!white}
\colorlet{cluster4}{blue!40!white}
\colorlet{cluster5}{cyan}
\colorlet{cluster6}{cyan!80!white}
\colorlet{cluster7}{cyan!60!white}
\colorlet{cluster8}{cyan!50!white}
\colorlet{cluster9}{cyan!40!white}
\colorlet{cluster10}{teal}
\colorlet{cluster11}{teal!80!white}
\colorlet{cluster12}{teal!60!white}
\colorlet{cluster13}{teal!50!white}
\colorlet{cluster14}{teal!40!white}

\colorlet{cluster0}{teal}
\colorlet{cluster1}{teal}
\colorlet{cluster2}{teal}
\colorlet{cluster3}{teal}
\colorlet{cluster4}{teal}
\colorlet{cluster5}{teal}
\colorlet{cluster6}{teal}
\colorlet{cluster7}{teal}
\colorlet{cluster8}{teal}
\colorlet{cluster9}{teal}
\colorlet{cluster10}{teal}
\colorlet{cluster11}{teal}
\colorlet{cluster12}{teal}
\colorlet{cluster13}{teal}
\colorlet{cluster14}{teal}

\colorlet{arima}{black}
\colorlet{price}{teal!50!black}
\colorlet{github}{green!75!black}
\colorlet{reddit}{red!75!black}
\colorlet{githubAndReddit}{blue!75!white}

\usepackage{blindtext}

\usepackage{xcolor,colortbl}  

\definecolor{darkpowderblue}{rgb}{0.0, 0.2, 0.6}

\definecolor{btc}{rgb}{0.93, 0.57, 0.13}
\definecolor{xmr}{rgb}{0.89, 0.26, 0.2}
\definecolor{etr}{rgb}{0.06, 0.05, 0.3}

\begin{document}

\maketitle

\begin{abstract}
 
Social media signals have been successfully used to develop large-scale predictive and anticipatory analytics. For example, forecasting stock market prices and influenza outbreaks. Recently, social data has been explored to forecast price fluctuations of cryptocurrencies, which are a novel disruptive technology with significant political and economic implications. In this paper we leverage and contrast the predictive power of social signals, specifically user behavior and communication patterns, from multiple social platforms GitHub and Reddit to forecast prices for three cyptocurrencies with high developer and community interest -- Bitcoin, Ethereum, and Monero. 
We evaluate the performance of neural network models that rely on long short-term memory units (LSTMs) trained on historical price data and social data against price-only LSTMs and baseline autoregressive integrated moving average (ARIMA) models, commonly used to predict stock prices. Our results not only demonstrate that social signals reduce error when forecasting daily coin price, but also show that the language used in comments within the official communities on Reddit (r/Bitcoin, r/Ethereum, and r/Monero) are the best predictors overall. We observe that models are more accurate in forecasting price one day ahead for Bitcoin (4\% root mean squared percent error) compared to Ethereum (7\%) and Monero (8\%).

\end{abstract}

\section{Introduction}
Cryptocurrencies, like Bitcoin, Ethereum, and Monero, are a new and disruptive technology that are often leveraged in highly volatile and fast-evolving environments.  
As with stocks and other securities, cryptocurrencies are bought, held, and traded. Unlike traditional currencies and stocks, these digital currencies rely on decentralized systems and cryptographic technologies, \eg blockchain ledgers, rather than a centralized institution, \eg banks. In this new paradigm, money is moved more quickly, independently, and often anonymously or semi-anonymously. As a result, the wide adoption and historic volatility of cryptocurrencies have significant political and economic implications. Price speculation, where traders buy securities in the hopes that they will quickly rise in price, often occurs in these highly volatile markets. Speculative trading typically occurs with little (or no) regard to the asset's fundamental value, but rather in regard to patterns in the the asset's price movement, rumor, or other suppositious data.

Signals from social media have been extensively used to predict real world events such as {election results}~\cite{tumasjan2010predicting,sang2012predicting,cameron2016can,dokoohaki2015predicting,wang2016boosting,khatua2015can}, movie sales~\cite{mishne2006predicting,asur2010predicting,tang2014information,abel2010analyzing}, 
{ protests}~\cite{maharjan2018towards},
{ public health events}~\cite{volkova2017forecasting,corley2010using,lamb2013separating,paul2014twitter,bodnar2013validating}, and { stock market activity}~\cite{bollen2011twitter,chen2014exploiting,makrehchi2013stock,oh2011investigating,mao2012correlating,martin2013predicting,porshnev2013machine,oliveira2013some,rao2012analyzing,zimbra2015stakeholder,li2016can,zhao2016correlating}. Ding \etal~\shortcite{ding2014using} introduced a deep neural network approach to predict the directionality of stock prices and the S\&P 500 index using signals from related news events and Tetlock~\cite{tetlock2007giving} highlighted the correlation between media pessimism and  market prices and volume. Bollen \etal~\shortcite{bollen2011twitter} predict relative differences in the daily Dow Jones industrial average using measures of collective mood states derived from Twitter activity and found the addition of some but not all possible states improved the predictive ability of their proposed models. Similar to Bollen~\etal, we analyze the benefit of including or excluding a range of social signals in our proposed models.

Like in stock markets and securities, cryptocurrency market activity has also been predicted using social signals. Previous work by Kim \etal~\shortcite{kim2016predicting} predicted the direction of price fluctuations for cryptocurrency coins utilizing social signals from online cryptocurrency forums. Other studies focus on predicting relative changes, \ie the return, in coin prices~\cite{rao2012analyzing,wang2017buzz}. For example, Wang and Verne~\shortcite{wang2017buzz} proposed a model to predict the return, \ie the change in price relative to the opening value. Phillips and Gorse~\shortcite{phillips2017predicting} predict the beginning and end of spikes in cryptocurrency prices, which they call  ``price bubbles'', using hidden Markov models that were previously used for the detection of influenza outbreaks. Although these previous works were able to predict certain changes in a cryptocurrency's price, they were not able to predict the actual price of the asset. In the present work we evaluate the benefit of incorporating a variety of social signals into models in order to forecast the {\it actual daily price high-values} of three popular cryptocurrencies.

\begin{figure}[t]
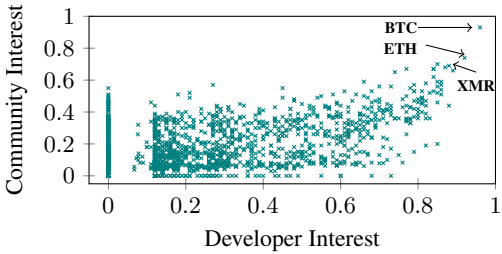
 
	\centering
	\include{figs/leidos_plot} 
	\vspace{-0.3in} 
	\caption{Coins plotted by developer and community interest. The present work will focus on three of the most popular: Bitcoin (BTC), Ethereum (ETH), and Monero (XMR)} 
	\label{fig:coin_clusters}  
\end{figure}

\begin{figure*}[t] 
	\centering 
	\small
	\input{figs/external_plus_activity_over_time_NCOMMENTS_BTC.tex} 
	\caption{Motivation: the alignment of price high (USD) and social interactions on Reddit (r/bitcoin) and GitHub (bitcoin/bitcoin).} 
	\label{fig:events}  
\end{figure*}
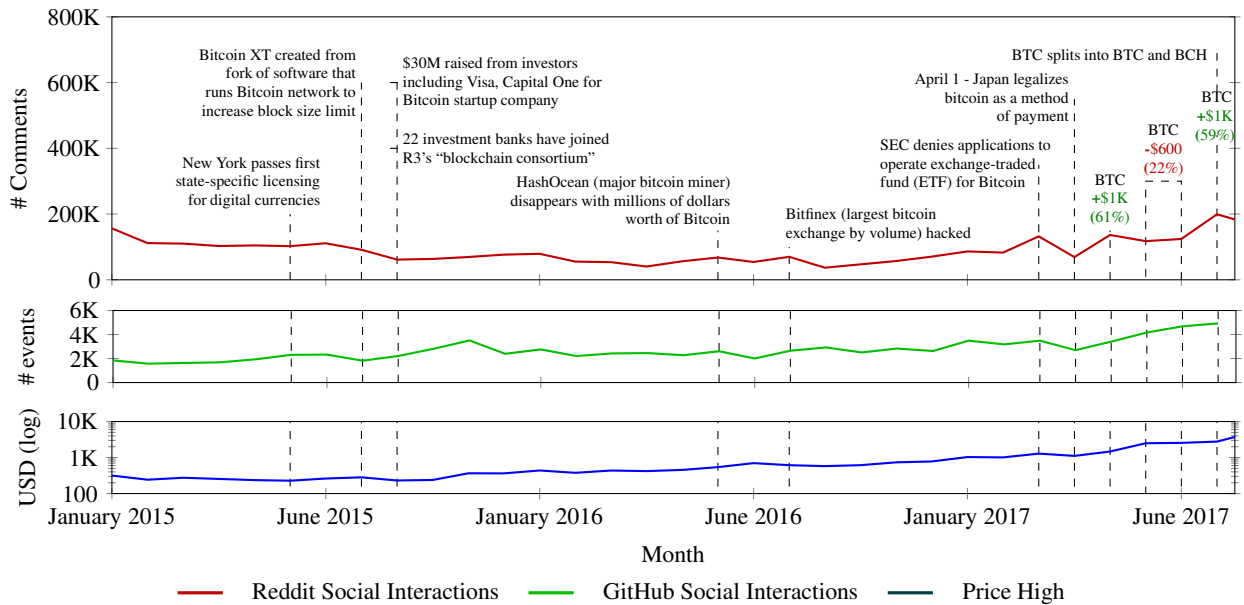

With the increasing use and reliance on these digital currencies, price fluctuation forecasting is an interesting yet difficult challenge. We address this problem by leveraging the predictive power of social signals from GitHub and Reddit to forecast immediate and near future prices of three popular cryptocurrencies: Bitcoin, Monero and Ethereum. In summary, our main contributions are:  
\begin{enumerate} 
	\item we develop neural network models that incorporate both social and price signals to generate forecasts of coin prices, 
	\item we present an in-depth analysis of model performance for coin price forecasts up to 3 days in advance for Bitcoin, Ethereum, and Monero when incorporating a variety of social signals and the relative improvement over models that rely solely on price history, and
	\item we report average performance of models for forecasts of coin price up to two weeks in advance.
\end{enumerate}

\subsection{Why Bitcoin, Ethereum, and Monero?}
A preliminary analysis of potential cryptocurrencies identified Bitcoin, Ethereum, and Monero as the top three cryptocurrencies in terms of both developer interest on GitHub and community interest on social media platforms. To determine which coins to focus on in this study, we collected data from CoinGecko\footnote{\url{https://www.coingecko.com/en/coins/all}} for 1,742 cryptocurrencies and performed an initial analysis of coins in terms of the developer interest (a measure of activity in public repositories on GitHub and Bitbucket) and community interest (a measure of discussions and popularity on social media) features released by CoinGecko. 
The key results of this analysis are illustrated in Figure~\ref{fig:coin_clusters}, which shows that Bitcoin, Ethereum, and Monero have the highest degree of both developer and community interest.

\section{Social and Financial Data Collection}

In this study, we focus on taking advantage of social data to forecast the price for three cryptocurrencies: Bitcoin (BTC), Ethereum (ETH), and Monero (XMR). Bitcoin is the first decentralized cryptocurrency and holds the highest market capitalization (market cap)\footnote{\label{cryptocompareFootnote}\url{https://www.cryptocompare.com/}}. Ethereum holds the second highest market cap and relies on the same blockchain technology that underpins Bitcoin. Monero, while still popular, holds a much lower market cap than Bitcoin or Ethereum but focuses on privacy; transactions are private (with the origin, destination, and amounts obfuscated) and untraceable (transactions cannot be linked to a particular cyber- or real-world identity).

Are social signals relevant to cryptocurrency prices? To gain an initial understanding of this question, we illustrate the alignment of price, social interactions, and real-world events related to Bitcoin in Figure~\ref{fig:events}.  

Historical price data (daily high, low, and price at market open and close) for each coin was collected from CryptoCompare\cref{cryptocompareFootnote}. This resulted in a price history from:
\begin{itemize}
	\item 2010/07/16 through 2018/05/21 for Bitcoin,
	\item 2015/08/06 through 2018/05/21 for Ethereum, 
	\item 2015/01/28 through 2018/05/21 for Monero. 
\end{itemize}
In addition to financial data, we collected publicly available data for two social platforms, GitHub and Reddit, from which we extracted social signals for each coin.

\paragraph{GitHub Dataset} GitHub is a collaborative software social network primarily used to develop and share novel technologies and software. As of 2017, 67M repositories (repos) used by 24M users and 1.5M organizations were hosted on the site.\footnote{\url{https://octoverse.github.com/}} We collected interactions with the main repo for each coin (\url {bitcoin/bitcoin}, \url{ethereum/go-ethereum}, and \url{monero-project/monero}) from a public subset of the GitHub archive.\footnote{\url{https://www.githubarchive.org/}} These interactions can be divided into two main categories of event types: 1) indications of user interest or repo \textit{popularity} such as watching or forking the repo and 2) \textit{direct contributions} when a user reviews or directly contributes to code, reports issues, comments on issues or participates in code reviews. 

\paragraph{Reddit Dataset} Similarly, we collected all posts and comments submitted to the official subreddits\footnote{Reddit posts and comments are publicly available via an archive hosted at \url{https://files.pushshift.io/reddit/}.} for each of the cryptocurrencies of interest (\url{r/bitcoin, r/ethereum}, and \url{r/monero}) across the three years (2015 -- 2017) for which we collected GitHub events data.
Reddit is a popular social news aggregator\footnote{The 18th most popular site globally and 5th most popular within the U.S. according to Alexa.com: \url{https://www.alexa.com/topsites}} that allows communities of users to share and discuss information, opinions, and entertainment media. Each of the subreddits has substantial traffic including 3.6K, 500, and 380 comments posted each day, on average, with community-sizes of 913K, 337K, and 137K subscribers (as of August 2018) for \url{r/bitcoin, r/ethereum}, and \url{r/monero}, respectively.

\section{Methodology}

In this section, we present the methodology used to evaluate the benefit of incorporating signals from social media into models that forecast the future price of a cryptocurrency. We trained and evaluated models that rely on 1) historical price alone, 2) historical price and each social signal, and 3) historical price and combinations of each of the social signals. 

\subsection{Forecasting Tasks}
We define the forecasting tasks as predicting the daily price high of each cryptocurrency of interest $j$ days in advance, $Y_{t_i+j}$, focusing on the immediate and near future $Y_{t_i+1}$, $Y_{t_i+2}$, $Y_{t_i+3}$ (1, 2, or 3 days in the future, respectively) using predictive signals from $k$ days in the immediate past $X_{[t_{i-k},t_{i}]}$ and evaluate any benefits that arise from incorporating up to two weeks of signal history by varying $k$ from 1 to 14 days. We consider models that incorporate social signals and historical price values versus those that consider only historical pricing as predictive signals to identify the performance gains (if any exist) when social signals are included. 
Models are trained and evaluated independently for each of the cryptocurrencies of interest.

\subsection{Social Signals} 
Along with the daily price high, we use a variety of signals of popularity, activity, and language used in discussions across two popular social media platforms: GitHub and Reddit. Here, we describe the social signals we extracted from social media activity related to the three coins of interest. 

\smallskip

\smallskip
\noindent
For the GitHub platform, we consider two types of social signals: 
\begin{itemize}
	\item $GH_{Pop}$ -- a vector representation of the daily totals for each \textbf{popularity} event: the Watch event, where users star a repo in order to receive notifications, and the Fork event, where users create a copy of the repo code. These event types provide a measure of how popular a given repo is among users who may or may not be direct contributors. 
	\item $GH_{All}$ -- a vector representation of the \textbf{overall activity}, \ie daily counts for each popularity (Watch and Fork) 
	and direct contribution event types (CommitComment, Issue, IssueComment, PullRequest, PullRequestReviewComment, and Push).
\end{itemize}

\smallskip
\noindent 
For the Reddit platform, we consider four types of social signals: 
\begin{itemize}
	\item $R_{Vol}$ -- the \textbf{volume} of comments posted each day, a signal for the size of discussion within the coin's official subreddit.
	\item $R_{Lang}$ -- the \textbf{language} used in comments 	represented as 10k-dimensional vectors of word-level daily-normalized statistics that focus on the most frequent unigrams. 
	\item $R_{Score}$ -- a vector representation of the quartiles of Reddit scores (\ie \# upvotes - \# downvotes) for comments posted each day, an indication of the range of \textbf{popularity} of comments for the given day. 
	\item $R_{Sent}$ -- quartiles for the subjectivity and polarity of comments each day  which provides a signal of the distribution of \textbf{sentiment} in discussions within the coin's official subreddit.
\end{itemize}

Before evaluating forecasting models,  we explore the relationships between coin-related social signals and their respective price high time-series. To do so, we first examine the correlation of social signals ($x$) and coin-price ($y$) over a sample of $N$ days for each of the three coins of interest using Pearson R correlation to examine the linear relationships between social signals and coin price. Pearson $R$ correlation ranges from -1 (perfectly inversely correlated) to 1 (perfectly correlated), where a score of 0 indicates linearly independent variables, \ie no correlation. Next, we consider non-linear correlations between signals ($x$) and price ($y$) using distance correlation~\cite{szekely2013energy,szekely2014partial,szekely2007measuring}: 
{\small
	\begin{displaymath}  
	dCorr(x,y) = \left\{\begin{array}{rr}
	\frac{dCov(x,y)}{\sqrt{dVar(x)dVar(y)} } ,&  dVar(x)dVar(y) > 0\\
	0,              &dVar(x)dVar(y) = 0
	\end{array}
	\right\}
	\end{displaymath} 
	where distance covariance ($dCov$) and distance variance ($dVar$) are defined as:
	\begin{displaymath}  
	dCov(x,y) = \sqrt{\frac{\sum_{k,l=1}^{n} A_{kl} B_{kl}}{n^2}},  
	~ dVar(x) = dCov(x,x)
	\end{displaymath} 
} 
\begin{displaymath} 
\textrm{and}  ~ A_{kl} = a_{kl} - \frac{1}{n}\sum_{l=1}^n a_{kl} - \frac{1}{n}\sum_{k=1}^n a_{kl} + \frac{1}{n^2}\sum_{k,l=1}^{n}a_{kl}.  
\end{displaymath}
$B_{kl}$ is defined similarly with $b_{kl}$ in place of $a_{kl}$ where $a_{kl}$ and $b_{kl}$ are Euclidean distance matrices of $x$ and $y$, respectively, defined as
\begin{displaymath} 
a_{kl} = |x_k-x_l|, ~~~ b_{kl} = |y_k-y_l|.
\end{displaymath}
Distance correlation varies from 0 to 1, where a distance correlation of 0 indicates independence of variables.
Finally, we also examine the variation of each signal ($x$) by the interquartile range (IQR) and standard deviation $\sigma$
for each of the signals of interest, identifying those that remain consistent over the time period of interest (and thus are potentially less informative).

\subsection{Forecasting Models} 
\begin{figure}[t] 
	\centering 
	\scalebox{1.2}{
		\input{figs/lstm_architecture_KDD} 
	} 
	\caption{Neural network model architecture.} 
	\label{fig:model_architecture}  
\end{figure}
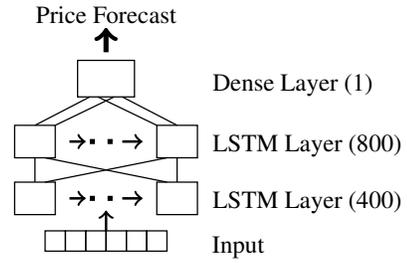

Neural network models with long short-term memory (LSTM) layers have previously been used to effectively forecast influenza dynamics from a combination of clinical and social media signals, outperforming models that did not incorporate the social signals~\cite{volkova2017forecasting}. LSTMs are a type of recurrent neural networks with built in memory cells that store information and can exploit long range context~\cite{hochreiter1997long}. These networks are surrounded by gating units that allow or prohibit the reset, read, and writing of such information. Inspired by the influenza dynamic models, we propose a neural network model, shown in Figure~\ref{fig:model_architecture}, that also utilizes LSTM layers. 

The proposed neural network architecture consists of a 400-dimensional LSTM layer followed by an 800-dimensional LSTM then a dense layer with a single unit. We do not employ an activation layer or activation function in the final dense output layer because we have defined each forecasting task as a regression task to predict numerical high values so an activation transformation is not needed. The optimizer used by the model is the ADAM optimizer (a parameter specific adaptive learning rate method)\cite{kingma2014adam} and we optimized performance by minimizing the mean squared error of a 20\% subset of the training data which is used as a validation set.  
To avoid overfitting, we employed early stopping callbacks with a maximum limit of 20 epochs. As a result, although models could have trained for up to 20 epochs, our models typically needed, on average, between five and six epochs only. The described architecture of the proposed model enables a relatively short training time for an expedient and scalable framework.

Signals used as input, e.g., social signals and historical price data, are fed to the network as a single concatenated vector. Before concatenation, each signal value is normalized to range between 0 and 1 using min-max normalization for the given feature across the entire dataset.  
Target price values ($T$) and signal input vectors ($I$) are defined as:
\begin{displaymath}
T(Y,j) = \{Y{j}, Y_{{j}+1}, ...Y_{|Y|}\}
\end{displaymath}
\begin{displaymath}
I(X,k) = \{ <X_{i-k},...,X_{i-1},X_i> for ~~~ i \in [k,|X|] ~\}
\end{displaymath}
where $X$ is a 2D array of the min-max normalized signals of interest, $Y$ is the historical price high time series, $j$ is the size of the forecasting window (\ie how many days in advance to predict the price), and $k$ is the number of days of signal history used.

\paragraph{Parameter Tuning} To identify our final model configuration, we performed a series of parameter tuning. We varied several of the model parameters: batch sizes of 16, 32, and 64; learning rates of 0.1, 0.01, 0.001, 0.0001, and 0.00001; and combinations of LSTM layers ranging from 10 to 400-dimensional layers followed by 20 to 800-dimensional layers. We found best performance when using a batch size of 16, learning rate of 0.001, and 400 and 800 units for the first and second LSTM layers, respectively.

\subsubsection{Baseline}
The autoregressive integrated moving average (ARIMA) model is a commonly used architecture when forecasting the price or return of stocks~\cite{pai2005hybrid,zhang2003time}. Essentially ARIMA models treat the future value of a variable (\eg price)  
as a linear combination of past values and errors.  We train a variety of ARIMA models and identify the top performing ARIMA model for each of our forecasting windows of interest  
to use as baselines. To identify these baseline models, we fix the size of the moving average window to 0, the difference order to 1 (to make the time series stationary), and perform autocorrelation analysis  
to identify a range of appropriate lag parameters. We found that models with a lag order of 0 achieved the best performance overall across coins.  
$R$ results for these baseline models were at least 0.95 ($p<0.001$) for Bitcoin, at least 0.92 ($p <0.001$) for Ethereum, and at least 0.91 ($p<0.001$) for Monero.

\subsection{Evaluation}  
To ensure performance comparison across identical train and test dates for all combinations of training and forecasting window sizes, the train and test periods were identified using the largest possible window sizes. We first restricted the dataset to the period for which we had data for all signals across all coins. Then, input vectors and target prices were formatted for the largest training (14) and forecasting (3) window sizes and split into 80\% for train and 20\% for test. This resulted in a training period from 2015/11/11 through 2017/04/18 (525 days) and a test period from 2017/05/04 through 2017/08/31 (120 days) used for all model configurations.

\ignore{
	\begin{algorithm}
		\caption{Formatting Input and Target Data}\label{alg:format}
		\begin{algorithmic}[1]
			\Input Price and Social Signal Input Vectors X, Price Time Series Y
			\Input Model Input
			\Procedure{MyProcedure}{}
			\State $\textit{stringlen} \gets \text{length of }\textit{string}$
			\State $i \gets \textit{patlen}$
			\BState \emph{top}:
			\If {$i > \textit{stringlen}$} \Return false
			\EndIf
			\State $j \gets \textit{patlen}$
			\BState \emph{loop}:
			\If {$\textit{string}(i) = \textit{path}(j)$}
			\State $j \gets j-1$.
			\State $i \gets i-1$.
			\State \textbf{goto} \emph{loop}.
			\State \textbf{close};
			\EndIf
			\State $i \gets i+\max(\textit{delta}_1(\textit{string}(i)),\textit{delta}_2(j))$.
			\State \textbf{goto} \emph{top}.
			\EndProcedure
		\end{algorithmic}
	\end{algorithm}
}

Model performance is evaluated using the following error measurements: root mean squared error (RMSE), mean squared percentage error (MSPE), mean absolute percentage error (MAPE), max absolute percentage error (MaxAPE), and root mean squared percentage error (RMSPE). For a set of $N$ predictions ($\hat{y}$) and true price values ($y$), MAPE, also known as mean absolute percentage deviation, is defined as:  
\begin{displaymath}
MAPE(\hat{y},y) =\frac{1}{N}\sum_{i=0}^{N}\frac{|\hat{y_i} - y_i|}{y_i} 
\end{displaymath}
\noindent maximum absolute percentage error is defined as:
\begin{displaymath}
MaxAPE(\hat{y},y) =\max \left( \frac{ |\hat{y_i} - y_i|}{y_i} \right)
\end{displaymath}
\noindent and mean squared percentage error is defined as:
\begin{displaymath}
MSPE(\hat{y},y) =  \frac{1}{N}\sum_{i=0}^{N}\frac{(\hat{y_i} - y_i)}{y_i}^2 
\end{displaymath}
Prices vary widely between coins and, for some, within coins over time. Therefore, we primarily report the performance metrics that consider percentage errors as they allow a relative comparison across the three coins of interest.

\section{Social Signal Analysis}
In this section we explore the relationships between daily price and signals of popularity and direct contributions to GitHub repositories and volume, sentiment, and popularity of cryptocurrency-related discussions on Reddit for our three coins of interest.  
To identify if an informative relationship exists, we examine person and distance correlation of social signals with coin price and the variance of all features to identify which signals to include in the battery of model configurations we consider in our ablation experiments.

In Table~\ref{tab:pearson_and_distance_correlation}, we see that price is correlated ($p<0.001$) with daily volume for each of the GitHub event types, although to varying degrees within and across coins. We also see that both Fork and Watch events are highly correlated for Bitcoin while Watch events are highly correlated and Fork events are moderately correlated for the other two coins.  
For the Reddit platform, we see that daily comment volume and comment scores are both correlated with daily price highs across all three coins of interest. While there is no significant linear correlation between subjectivity and polarity of the daily batch of comments and price highs for Bitcoin, these features have varying levels of correlation across the remaining two coins of interest. As we saw with GitHub features, there is some variation across coins of interest in the relationships between the various social signals and price timeseries.

\begin{table}[t] 
	\caption{Pearson $R$ correlation and distance correlation ($DC$) of daily price and social signals.  
		Pearson results are significant ($p<0.001$) unless indicated 
		with a dash `---'  ($p\ge 0.05$).} 
	\centering
	\small 
	\setlength\tabcolsep{2pt} 
	\begin{tabular}{llccccccccc}
		&
		
		&\multicolumn{2}{c}{Bitcoin}&& \multicolumn{2}{c}{Ethereum}&& \multicolumn{2}{c}{Monero}\\
		&Social Signal &\tiny $R$ & \tiny $DC$ &&\tiny $R$ & \tiny $DC$ &&\tiny $R$ & \tiny $DC$ \\
		\hline
		\multirow{8}{0.5em}{\rotatebox{90}{GitHub}}&

		Watch&{\cellcolor{teal!87}0.87 } & {\cellcolor{teal!86} 	0.86}
		& &
		{\cellcolor{teal!68 }0.68 } & {\cellcolor{teal!73} 	0.73	}
		&	 &
		{\cellcolor{teal!72 }0.72 } & {\cellcolor{teal!68} 	0.68	}
		\\
		
		& 
		Fork&{\cellcolor{teal!75}0.75 } & {\cellcolor{teal!72} 	0.72}
		& &
		{\cellcolor{teal!40 }0.40 } & {\cellcolor{teal!38} 	0.38	}
		&&
		{\cellcolor{teal!41 }0.41 } & {\cellcolor{teal!48} 	0.48	}
		\\
		
		& 
		Issues&{\cellcolor{teal!5 }0.05 } & {\cellcolor{teal!9} 	0.09}
		& &
		{\cellcolor{teal!22 }0.22 } & {\cellcolor{teal!5} 	0.05	}
		& &
		{\cellcolor{teal!00 }0.00 } & {\cellcolor{teal!36} 	0.36	}
		\\
		
		& 
		IssueComment&{\cellcolor{teal!13 }0.13 } & {\cellcolor{teal!14} 	0.14	}
		&&
		{\cellcolor{teal!36}0.36 } & {\cellcolor{teal!29} 	0.29	}
		& &
		{\cellcolor{teal!25 }0.25} & {\cellcolor{teal!54} 0.54	}
		\\
		
		& 
		Push&{\cellcolor{teal!06 }0.06 } & {\cellcolor{teal!9} 0.09}
		& &
		{\cellcolor{teal!06}0.06 } & {\cellcolor{teal!17} 	0.17	}
		& &
		{\cellcolor{teal!07 }0.07 } & {\cellcolor{teal!	11	} 	0.11	}
		\\
		
		& 
		CommitComment&{\cellcolor{teal!06 }0.06 } & {\cellcolor{teal!	9	} 	0.09	}
		& &
		{\cellcolor{teal!04 }0.04 } & {\cellcolor{teal!	12	} 	0.12	}
		&&
		{\cellcolor{teal!08 }0.08 } & {\cellcolor{teal!	8	} 	0.08	}
		\\
		
		& 
		PullRequest (PR)&{\cellcolor{teal!18}0.18} &  {\cellcolor{teal!	20	} 	0.20	}
		&&
		{\cellcolor{teal!14 }0.14 } &  {\cellcolor{teal!	22	} 	0.22	}
		& &
		{\cellcolor{teal!15 }0.15 } & {\cellcolor{teal!	21	} 	0.21	}
		\\
		
		& 
		PRReviewComment&{\cellcolor{teal!39 }0.39} &{\cellcolor{teal!	37	} 	0.37	}
		&&
		{\cellcolor{teal!29 }0.29 } &{\cellcolor{teal!	36	} 	0.36	}
		&&
		{\cellcolor{teal!22 }0.22 } &{\cellcolor{teal!	44	} 	0.44	}
		\\

		\hline

		\multirow{4}{0.5em}{\rotatebox{90}{Reddit}}&  
		Comment Volume  & {\cellcolor{teal!58}   0.58 	} & {\cellcolor{teal!	62	} 	0.62	}
		&  
		& {\cellcolor{teal!48}  0.48 }  & {\cellcolor{teal!	51	} 	0.51	}
		& 
		&  {\cellcolor{teal!67} 0.67 }  & {\cellcolor{teal!	78	} 	0.78	}
		\\
		&  
		Subjectivity  &    --- 	& {\cellcolor{teal!	0	} 	0.00	}
		&    
		&   ---				& {\cellcolor{teal!	5	} 	0.05	}
		&   
		&  {\cellcolor{teal!16} 0.16 }  & {\cellcolor{teal!	25	} 	0.25	}
		\\
		&  
		Polarity  &   --- 		& {\cellcolor{teal!	2	} 	0.02	}
		&  
		&  {\cellcolor{teal!13} 0.13 }  & {\cellcolor{teal!	15	} 	0.15	}
		&  
		& {\cellcolor{teal!31}  0.31 }  & {\cellcolor{teal!	43	} 	0.43	} 
		\\
		&  
		Score &  {\cellcolor{teal!34} 0.34} & {\cellcolor{teal!	37	} 	0.37	}
		&  
		& {\cellcolor{teal!47}  0.47 } & {\cellcolor{teal!	59	} 	0.59	}
		&  
		&  {\cellcolor{teal!69} 0.69 } &{\cellcolor{teal!	79	} 	0.79	}
		\\
		\hline 
		\\
	\end{tabular}
	\label{tab:pearson_and_distance_correlation}
\end{table}

\begin{table}[t] 
	\caption{Standard deviations ($\sigma$) and Inter-Quartile Range (IQR) of daily price and social signals.} 
	\centering
	\small 
	\setlength\tabcolsep{2pt} 
	\begin{tabular}{lrrrrrr} 
		\hline
		&	\multicolumn{2}{c}{Bitcoin}			&	\multicolumn{2}{c}{Ethereum}			&	\multicolumn{2}{c}{Monero}			\\
		Signal	&	$\sigma$	& \tiny	$IQR$	&	$\sigma$	&\tiny	$IQR$	&	$\sigma$	&\tiny	$IQR$	\\
		\hline
		\tiny	Price High	&	868.00	&	626.85	&	99.44	&	34.67	&	19.71	&	11.96	\\
		\hline													
		\tiny	Watch	&	10.34	&	7.00	&	2.09	&	2.00	&	1.60	&	1.00	\\
		\tiny	Fork	&	4.55	&	4.00	&	0.66	&	0.00	&	0.94	&	1.00	\\
		\tiny	Issues	&	3.34	&	3.00	&	27.42	&	1.00	&	2.47	&	2.00	\\
		\tiny	IssueComment	&	24.98	&	34.00	&	7.93	&	7.00	&	9.51	&	10.00	\\
		\tiny	Push	&	3.57	&	5.00	&	1.06	&	0.00	&	3.80	&	1.00	\\
		\tiny	CommitComment	&	0.90	&	0.00	&	0.18	&	0.00	&	0.26	&	0.00	\\
		\tiny	PullRequest (PR)	&	5.92	&	8.00	&	1.14	&	0.00	&	5.15	&	4.00	\\
		\tiny	PRReviewComment	&	13.25	&	15.00	&	2.57	&	0.00	&	5.50	&	1.00	\\
		\hline													
		\tiny	Comment Volume	&	1990.43	&	1981.00	&	695.86	&	369.00	&	376.34	&	466.00	\\
		\tiny	Score	&	242.97	&	129.00	&	80.00	&	47.00	&	18.34	&	19.00	\\
		\tiny	Subjectivity	&	0.00	&	0.00	&	0.00	&	0.00	&	0.13	&	0.00	\\
		\tiny	Polarity	&	0.01	&	0.00	&	0.06	&	0.00	&	0.23	&	0.25	\\
		\hline

	\end{tabular}
	\label{tab:variation_in_signals}
\end{table}

When we consider how the signals vary within themselves in Table~\ref{tab:variation_in_signals}, we find simaliar patterns across and wihtin coins. The most highly correlated social signals have larger, relative to other signals, variation as summarized by the standard deviations and inter-quartile ranges of the signal vectors. We find that the subjectivity and polarity signals from Reddit comments linked to Bitcoin and Ethereum that held no significant correlation with price also show little to no variation. However, we see they vary slightly to moderately for the third coin, Monero. As each of our social signals indicate a relationship with price for at least one of our coins of interest, we include all GitHub and Reddit social signals as well as combinations of signals from both the GitHub and Reddit platforms in our ablation experiments and highlight the results in the following section.

\section{Forecasting Results}
In this section, we describe the performance of models that rely on historical price alone, historical price and each signal from GitHub or Reddit, and historical price and combinations of each GitHub and Reddit signal. In particular, we highlight models which incorporated social signals that achieved high performance relative to the baseline ARIMA models and LSTMs that relied on historical price alone.

\begin{figure}[t] 
	\centering 
	\small 	
	\input{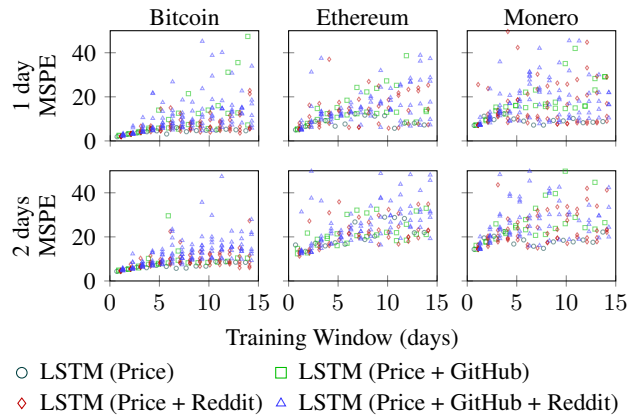}
	\caption{MSPE results for neural network models for each coin plotted by varying training window sizes for the first two forecasting windows, \ie when predicting one or two days in advance. Jitter has been added to x-axis for enhanced readability.}
	\label{fig:varying_training_window} 
\end{figure}

First, we explored the benefit of increased signal history with a comparison of model performance using signals from one to fourteen days in the past in Figure~\ref{fig:varying_training_window}. We plot MSPE as a function of training window size, \ie the number of days of signal history to rely on, for neural network (LSTM) models that rely on each of the combinations of predictive signals when forecasting price one to two days in advance. {\it Interestingly, we see that models with the smallest window size (1) achieved the best performance.}  Therefore, we use a signal history window size of 1 day for subsequent model evaluation.

Next we focused on identifying model configurations that achieve the best performance to evaluate the benefit of incorporating social signals from a variety of platforms and the benefit of a variety of combinations of social signals. To do so, we perform an ablation study where we compare models that incorporate each combination of price and social signals with models (LSTM and ARIMA baselines) that relied solely on historical prices. To identify the best, overall model, we then averaged percentage errors across the three coins and ranked social signal-infused models with the baseline ARIMA and neural network models that did not incorporate social signals. 

We present a summary of the averaged error for the top performing models in Table~\ref{tab:meanRMSPE_3days}, ordered by the mean of mean RMSPE over the three forecasting tasks. {\it The top performing model is the proposed LSTM that relies on price history and $R_{Lang}$, the representation of the language used in comments within the official subreddit.} Here, we see that LSTM models that incorporate social signals outperform the price only baselines when averaged across the three coins and forecasting windows. If we only consider the immediate forecasting window of one day in advance, the proposed LSTM model that relies solely on price history outperforms the others.

We then consider performance for each coin individually in Table~\ref{tab:MAPE_results}. Here we see that, in most cases, neural network models that incorporated social signals slightly outperformed both ARIMA baselines and neural network models that relied solely on price history, but these results are not statistically significant. However, we see that our proposed LSTM neural network architecture, and especially neural network models that incorporate social signals, minimize the maximum absolute percentage error (MaxAPE). That is, in worst-case prediction performance, we see a much lower error rate for our proposed models that rely on social signals alongside price history. We see the best performance is achieved when models forecast the price one day in the future (FW = 1 Day); unsurprisingly, it is easiest to predict the next day's price using signals from the day before.

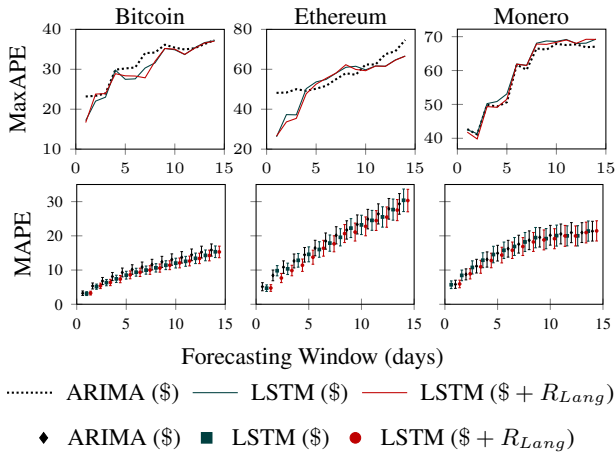
\begin{figure}[h!] 
	\centering 
	\small  
	\input{figs/performance_by_FW_twoweeks_plot.tex}
	\caption{MaxAPE (above) and MAPE with 95\% confidence intervals (below) for each coin plotted by varying forecasting window sizes (1 to 14 days in advance). Jitter has been added to x-axis of MAPE plots for enhanced readability.} 
	\label{fig:two_week_forecasting_errors} 
\end{figure}

If we expand the range of forecasting windows beyond the near and immediate future, we see that, again, the LSTM model that incorporates the $R_{Lang}$ social signal alongside historical price has the best performance across the three coins. Table~\ref{tab:meanRMSPE_14days} presents the RMSPE of top performing models when forecasting price up to two weeks in advance.  Figure~\ref{fig:two_week_forecasting_errors} illustrates how this top performing model, performs similarly to price-only LSTM and ARIMA models, on average, outperforms the price-only models in respect to the worst-case prediction errors (MaxAPE).

\begin{table}[t]
	\caption{RMSPE averaged over the three coins of interest (Bitcoin, Ethereum, and Monero) when predicting price up to 3 days in advance for ARIMA baseline and top performing neural network (LSTM) models. Lowest values highlighted in bold.}
	\small
	\centering
	\setlength\tabcolsep{4.5pt} 
	\begin{tabular}{ll|rrr|r}
		\hline
		&&\multicolumn{4}{c}{Forecasting Window (days)}\\
		Model&Signal &  1 &  2 &  3 &  mean \\
		\hline
		LSTM &            \tiny $\$+R_{Lang}$ & 
		6.70 &     \bf       9.88 &      \bf     12.06 &  \bf     9.55 \\
		LSTM &   \tiny $\$+GH_{Pop}+R_{Lang}$ &
		6.64 &            9.99 &           12.40 &       9.68 \\
		LSTM &             \tiny $\$+R_{Vol}$ &
		6.78 &            9.98 &           12.48 &       9.75 \\
		LSTM &                     \tiny $\$$ & 
		\bf 6.60 &           10.46 &           12.32 &       9.79 \\
		ARIMA &                     \tiny $\$$ & 
		7.30 &           10.56 &           13.10 &      10.32 \\
		\hline 
	\end{tabular}
	\label{tab:meanRMSPE_3days}
\end{table}

\begin{table*}[t]
	\caption{MAPE, RMSPE, and MaxAPE results for baseline ARIMA, neural network models that rely solely on historical price, and the top performing social signal enhanced neural network models, as identified during ablation experiments.  
		Results for neural network models (LSTM) reported used the top performing training window size of 1 day. Lowest error rates are highlighted in bold. } 
	\centering
	\small 
	\setlength\tabcolsep{2pt} 
	
	\begin{tabular}{lrlrrr|rrr|rrr} 
		\hline
		& & & 
		\multicolumn{3}{c|}{Bitcoin}  & \multicolumn{3}{c|}{Ethereum} & 
		\multicolumn{3}{c}{Monero}    \\
		
		\textcolor{white}{FI} & Model &  Signals &   \multicolumn{1}{c}{\tiny MAPE} & \multicolumn{1}{c}{\tiny  RMSPE} &  \multicolumn{1}{c|}{\tiny MaxAPE} &  \multicolumn{1}{c}{\tiny MAPE} & \multicolumn{1}{c}{\tiny  RMSPE} &  \multicolumn{1}{c|}{\tiny MaxAPE} &   \multicolumn{1}{c}{\tiny MAPE} & \multicolumn{1}{c}{\tiny  RMSPE} &  \multicolumn{1}{c}{\tiny MaxAPE}\\
		\hline
		
		\multirow{5}{0em}{\rotatebox{90}{\$ in 1 day}}
		&\tiny ARIMA & $\$$  & 3.24 $\pm$ 0.65
		& 4.83 $\pm$ 3.35
		& 23.19 
		& 5.13 $\pm$ 1.18
		& 8.29 $\pm$ 6.50
		& 48.19 
		& 5.84 $\pm$ 1.19
		& 8.78 $\pm$ 6.26
		& 42.71 
		\\
		&\tiny LSTM & $\$$  & \bf 3.11 $\pm$ 0.55
		& \bf 4.35 $\pm$ 2.69
		& 17.16 
		& \bf 4.71 $\pm$ 0.96
		& \bf 7.08 $\pm$ 4.48
		& \bf 26.18 
		& \bf 5.71 $\pm$ 1.12
		& 8.39 $\pm$ 5.96
		& 42.53 
		\\
		&\tiny LSTM & $\$+R_{Lang}$  & 3.26 $\pm$ 0.55
		& 4.45 $\pm$ 2.69
		& \bf 16.66 
		& 4.76 $\pm$ 0.97
		& 7.16 $\pm$ 4.54
		& 26.53 
		& 5.94 $\pm$ 1.11
		& 8.49 $\pm$ 5.85
		& \bf 41.83 
		\\
		&\tiny LSTM & $\$+GH_{Pop}+R_{Lang}$  & 3.19 $\pm$ 0.56
		& 4.44 $\pm$ 2.68
		& 16.77 
		& 4.73 $\pm$ 0.97
		& 7.10 $\pm$ 4.50
		& 26.35 
		& \bf 5.71 $\pm$ 1.12
		& \bf 8.38 $\pm$ 5.97
		& 42.70 
		\\
		&\tiny LSTM & $\$+R_{Vol}$  & 3.35 $\pm$ 0.55
		& 4.52 $\pm$ 2.71
		& 16.49 
		& 5.06 $\pm$ 0.99
		& 7.44 $\pm$ 4.68
		& 27.36 
		& 5.69 $\pm$ 1.12
		& 8.38 $\pm$ 5.99
		& 42.72 
		\\
		\hline
		\multirow{5}{0em}{\rotatebox{90}{\$ in 2 days}}&\tiny ARIMA & $\$$  & 5.35 $\pm$ 0.87
		& 7.18 $\pm$ 4.13
		& 23.35 
		& 8.44 $\pm$ 1.56
		& 12.05 $\pm$ 7.83
		& 48.35 
		& 8.74 $\pm$ 1.61
		& 12.44 $\pm$ 7.56
		& 41.13 
		\\
		&\tiny LSTM & $\$$  & 5.14 $\pm$ 0.76
		& 6.61 $\pm$ 3.68
		& 22.03 
		& 9.81 $\pm$ 1.49
		& 12.78 $\pm$ 6.84
		& 37.30 
		& 8.44 $\pm$ 1.55
		& 11.99 $\pm$ 7.40
		& 41.35 
		\\
		&\tiny LSTM & $\$+R_{Lang}$  & 5.38 $\pm$ 0.79
		& 6.90 $\pm$ 3.82
		& 23.79 
		& \bf 7.67 $\pm$ 1.34
		& \bf 10.61 $\pm$ 6.07
		& \bf 33.64 
		& 8.90 $\pm$ 1.50
		& 12.12 $\pm$ 7.26
		& \bf 39.80 
		\\
		&\tiny LSTM & $\$+GH_{Pop}+R_{Lang}$  & 5.40 $\pm$ 0.78
		& 6.89 $\pm$ 3.78
		& 23.23 
		& 7.87 $\pm$ 1.43
		& 11.11 $\pm$ 6.33
		& 35.01 
		& \bf 8.32 $\pm$ 1.56
		& \bf 11.95 $\pm$ 7.48
		& 41.73 
		\\
		&\tiny LSTM & $\$+R_{Vol}$  & \bf 5.11 $\pm$ 0.76
		& \bf 6.59 $\pm$ 3.67
		& \bf 21.78 
		& 8.11 $\pm$ 1.44
		& 11.34 $\pm$ 6.38
		& 35.31 
		& 8.44 $\pm$ 1.56
		& 12.02 $\pm$ 7.48
		& 41.80 
		\\
		\hline
		\multirow{5}{0em}{\rotatebox{90}{\$ in 3 days}}&\tiny ARIMA & $\$$  & 6.80 $\pm$ 1.03
		& 8.85 $\pm$ 4.72
		& 23.94 
		& 10.85 $\pm$ 1.83
		& 14.81 $\pm$ 8.74
		& 50.03 
		& 11.14 $\pm$ 1.99
		& 15.64 $\pm$ 9.17
		& 49.89 
		\\
		&\tiny LSTM & $\$$  & \bf 6.33 $\pm$ 0.92
		& \bf 8.08 $\pm$ 4.29
		& \bf 23.01 
		& 10.36 $\pm$ 1.66
		& 13.79 $\pm$ 7.35
		& 37.17 
		& \bf 10.77 $\pm$ 1.92
		& \bf 15.07 $\pm$ 8.90
		& 50.23 
		\\
		&\tiny LSTM & $\$+R_{Lang}$  & 6.37 $\pm$ 0.92
		& 8.12 $\pm$ 4.31
		& 23.92 
		& 9.86 $\pm$ 1.53
		& 12.96 $\pm$ 6.90
		& 35.41 
		& 10.96 $\pm$ 1.90
		& 15.10 $\pm$ 8.85
		& \bf 49.36 
		\\
		&\tiny LSTM & $\$+GH_{Pop}+R_{Lang}$  & 6.36 $\pm$ 0.92
		& 8.13 $\pm$ 4.30
		& 23.13 
		& \bf 9.84 $\pm$ 1.53
		& \bf 12.92 $\pm$ 6.87
		& \bf 35.31 
		& 11.77 $\pm$ 2.02
		& 16.16 $\pm$ 9.28
		& 52.58 
		\\
		&\tiny LSTM & $\$+R_{Vol}$  & 6.36 $\pm$ 0.92
		& 8.13 $\pm$ 4.28
		& 23.08 
		& 10.38 $\pm$ 1.67
		& 13.84 $\pm$ 7.37
		& 37.31 
		& 10.97 $\pm$ 1.99
		& 15.46 $\pm$ 9.07
		& 51.53 
		\\
		\hline

		\hline

	\end{tabular}

	\label{tab:MAPE_results}
\end{table*}

\begin{table*}[t]
	\caption{RMSPE averaged over the three coins of interest (Bitcoin, Ethereum, and Monero) when predicting price up to 14 days in advance for ARIMA baseline and top performing neural network (LSTM) models.}
	\small
	\setlength\tabcolsep{3pt} 
	\begin{tabular}{ll|rrrrrrrrrrrrrr|r}
		\hline 
		
		&	& \multicolumn{14}{c|}{Forecasting Window (days)}\\
		Model &  Signals &  1 &  2 &  3 &  4 &  5 &  6 &  7 &  8 &  9 &  10 &  11 &  12 &  13 &  14 &  mean \\
		\hline
		
		\tiny LSTM &            
		$\$+R_{Lang}$ &            6.70 &   \bf         9.88 &     \bf      12.06 &     \bf      13.92 &      \bf     15.83 &           17.57 &         \bf  18.72 &           20.45 &          \bf  21.22 &      \bf      22.25 &        \bf    23.38 &          \bf  23.87 &            25.23 &            26.67 &      \bf 18.41 \\
		\tiny LSTM &                     
		$\$$ &     \bf       6.60 &           10.46 &           12.32 &           14.63 &           16.35 &           17.38 &           18.74 &           20.08 &           21.92 &            22.44 &            23.43 &            23.92 &     \bf       24.96 &            26.71 &      18.57 \\
		\tiny LSTM &            
		$\$+GH_{Pop}$ &            6.80 &           10.27 &           12.76 &           14.50 &           16.19 &           \bf 17.36 &           18.75 &     \bf      19.97 &           21.40 &            23.13 &            23.76 &            24.26 &            25.66 &            27.32 &      18.72 \\
		\tiny LSTM &   
		$\$+GH_{Pop}+R_{Lang}$ &            6.64 &            9.99 &           12.40 &           14.19 &           16.17 &           17.66 &           19.14 &           20.56 &           21.62 &            23.16 &            24.01 &            24.90 &            26.11 &            26.67 &      18.80 \\
		\tiny ARIMA &                     
		$\$$ &            7.30 &           10.56 &           13.10 &           15.07 &           16.62 &           18.15 &           19.62 &           20.95 &           22.17 &            23.17 &            23.99 &            24.88 &            25.58 &        \bf    26.32 &      19.11 \\

		\hline
	\end{tabular}
	\label{tab:meanRMSPE_14days}
\end{table*}

\begin{figure*}[t!] 
	\centering 
	\small
	\input{figs/price_and_reddit_language_predictions.tex} 
	\vspace{-0.1in}
	\caption{Predictions from the overall top performing neural network model (incorporating price and $R_{Lang}$, language used in comments posted to the official subreddit communities). Solid gray lines represent the true price high values for each day across the test period (2017/05/04-2017/08/31) and colored lines with markers (representing individual predictions) plot predictions.} 
	\label{fig:prediction_timeseries} 
\end{figure*}
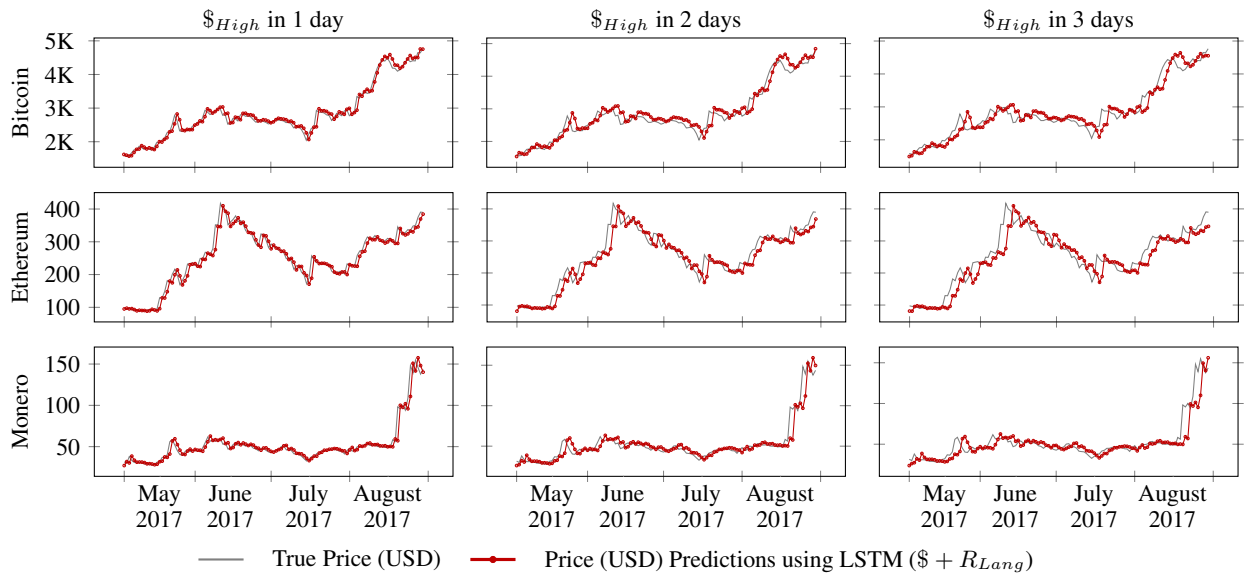

We illustrate the actual predictions of the top performing social signal enhanced neural network model against the true price values over the test period in Fig.~\ref{fig:prediction_timeseries}. This allows us to visualize not only the performance of each model relative to the true price by day but also relative to the trends in true price values across coins. As we saw in Table~\ref{tab:variation_in_signals}, the range of true price values differs greatly by coin. While Bitcoin ranges from \$1000 to \$5000, Ethereum ranges from \$100 - \$400 and Monero ranges between \$30 and \$50 for the majority of the test period with a spike to between \$100 and \$150 in the final 10 days of the test period. {\it The differences in the variance of true price parallel the differences in model performance.} These results indicate that variations in price or the range of price values may heavily affect predictability of coin price; coins with lower price values or variance are more difficult to predict.

\section{Discussion}
\begin{figure}[h!] 
	\centering 
	\small
	\input{figs/best_PplusSM_models_predictions/mape_by_genesis_block_date.tex} 
	
	
	\input{figs/best_PplusSM_models_predictions/mape_by_marketcap.tex}


	\input{figs/best_PplusSM_models_predictions/mape_by_std.tex}

	\begin{tikzpicture} 
	\begin{axis}[%
	hide axis,
	xmin=10,xmax=50,ymin=0,ymax=0.4,
	legend style={draw=none,legend cell align=left,legend columns = -1, column sep = 3mm}
	] 
	\addlegendimage{arima, only marks, mark size=3,mark=+,thick}; 
	\addlegendentry{ARIMA}; 
	\addlegendimage{price, only marks,mark size=3,mark=x,thick}; 
	\addlegendentry{LSTM (Price)}; 
	\addlegendimage{github, only marks, mark=o,thick}; 
	\addlegendentry{LSTM (Price+GitHub)}; 
	\end{axis}
	\end{tikzpicture} 
	
	
	\begin{tikzpicture} 
	\begin{axis}[%
	hide axis,
	xmin=10,xmax=50,ymin=0,ymax=0.4,
	legend style={draw=none,legend cell align=left,legend columns = -1, column sep = 3mm}
	]  
	\addlegendimage{reddit, only marks, mark=o,thick};  
	\addlegendentry{LSTM (Price+Reddit)}; 
	\addlegendimage{githubAndReddit, only marks, mark=o,thick}; 
	\addlegendentry{LSTM (Price+GitHub+Reddit)}; 
	\end{axis}
	\end{tikzpicture} 
	\vspace{-0.15in}
	\caption{MaxAPE as a function of lifetime (\# days between mining of genesis block and end of test period), as a function of market cap (in billion USD), and standard deviations of coin price high (USD) for ARIMA and LSTM price only  and top performing price and social signal LSTM models.}
	\label{fig:mape_by_marketcap} 
\end{figure}
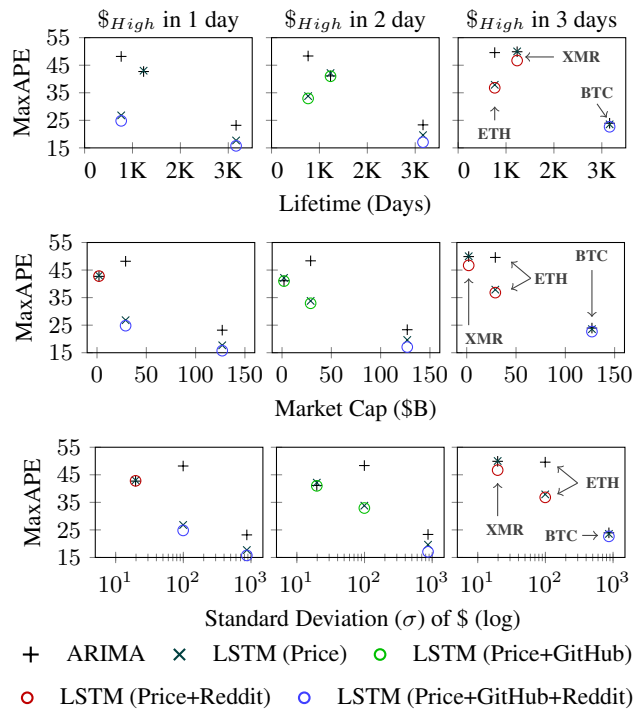

In each of our analyses it is apparent that models perform best when forecasting the price for Bitcoin then Ethereum and worst when forecasting for Monero. In Fig.~\ref{fig:prediction_timeseries}, we saw evidence that this may be attributable to variations in true price but we hypothesize it may also be tied to the size of activity on social media or market cap -- Monero has both the smallest social activity from which we drew social signals (GitHub event and comment counts noted above when we described data collection) and the lowest market cap (as is shown in Fig.~\ref{fig:mape_by_marketcap}). As we saw in Fig.~\ref{fig:prediction_timeseries}, Monero also had the smallest range of price values. This suggests that activity volumes may have a direct effect on cryptocurrency prices, or at the least, performance of models that forecast coin price.

As we saw above, the best model performance is achieved for forecast windows of 1 day in advance, an intuitive result as this is an easier task than longer windows of 2 and 3 days. However, we also find better performance across all forecasting windows for Bitcoin. Bitcoin is both the oldest and most established coin with the largest market cap. Within our three coins of interest, Ethereum has the next highest market cap but Monero has the next longest lifetime. As a result, we plot MAPE as a function of lifetime and as a function of market cap\footnote{Market cap for each coin collected from CryptoCompare.com on September 4th 2018.} in Fig.~\ref{fig:mape_by_marketcap} to explore whether one of these characteristics may have an effect on price values or model performance. We calculate lifetime as the number of days from when the genesis block was mined (\ie the beginning of the ``ledger'' of transactions) to the end of the testing period (2017/08/31). 

When we examine Fig.~\ref{fig:mape_by_marketcap},  we find that MaxAPE tends to decrease as market cap increases among top performing models but we do not find the same pattern when we plotted MAPE by lifetime. It is important to note that the sample size of coins is small (N=3) and these patterns may not hold among a larger sample of coins or among a set of coins with more variation in lifetimes or market cap. However, these results combined with the increased performance of models that rely on indications of popularity (event counts, comment volume, etc.) suggest that the popularity of a coin affects the performance of predictive models that forecast coin prices. Intuitively, popularity should affect price -- a coin that no one knows about or that is less popular probably has a lower price.

There are several potential impacts of identifying models that reliably forecast the actual prices of cryptocurrencies, beyond the obvious use of identifying which coin to purchase and when. One such use would be as indications of irregular and potentially fraudulent or deceptive activity. For example, a ``pump and dump'' where a group artificially inflates the price of a coin with excessive transactions and then sells to unsuspecting speculators outside the group at the peak. Although illegal in stocks and securities, this behavior is not yet regulated in cryptocurrencies and is openly advertised~\cite{fuscaldo2018pump,town2018pump}. Similarly, fabricated or misleading news stories and social media postings have been used to artifically inflate or decrease coin prices.  Say there is a model that reliably predicts the price of a cryptocurrency and it begins to fail at an unexpected magnitude, could this be used as an indication of ``pump and dump'' activity or other manipulation?

\section{Conclusions and Future Work}
In the current work, we have presented models that incorporate social signals for improved forecasts of cryptocurrency price values. Rather than predicting the returns (relative price difference) or the direction of fluctuations (increase versus decrease in price) -- both are much easier tasks compared to our task formulation, our models predict the daily price high in USD.  We focused on social signals from the language, volume, and sentiment of discussions on Reddit and indications of popularity and direct contribution activity on GitHub.  Our analysis of model performance and comparisons to baselines that rely solely on price history have identified the benefit of social signals. 
Although performance improvements, on average, are not statistically significant, in a task such as price value prediction in a market as volatile as cryptocurrencies, even a modest performance improvement can have a notable impact. More so, the minimization of worst-case error (MaxAPE) when using our proposed social signal-infused model is of significant benefit.

With the speed and volatility of cryptocurrency markets, it may be more valuable to model price in granularities of hours or minutes rather than days. However, a limitation of this approach is the accessibility of such fine-grained pricing data.  As the price data available was provided in daily increments, we presented the results of models for predictions of more coarse-grained daily price high-values. The proposed neural network models and framework of social signal extraction could be easily adapted to a more fine-grained approach.

Due to data collection (API) constraints, we performed a similar analysis of models using Twitter social signals for a subset of the time period of interest. Results were inconclusive but future work will also consider such social signals from Twitter and other related platforms alongside GitHub and Reddit.  
We are also in the process of collecting data to cover not only the time period considered in the current work but also a second dataset that considers the periods of historic spikes and dips in Bitcoin pricing, \ie the period covering the significant spike in the price of Bitcoin at the end of 2017 and steep drops that occurred over the first few months of 2018.

Our analysis and discussion have also identified several other avenues of future work. One such avenue is the potential to generalize the top performing models for the most popular coin or coins, \eg Bitcoin, to forecast price values for other coins, such as new or less popular coins. Can we combine the social signals for Bitcoin or cryptocurrencies in general with a coin's historical price features to predict coins that do not have a sufficient volume of social activity? Another direction for future work is the adaption of these models for use in anomaly detection --- can models be adapted to identify coins that may be compromised (\eg through targeted deception in news stories and social media discussions or fraudulent behavior like ``pump and dump'' schemes)? Finally, a third avenue is an expanded exploration into the trends identified as potential drivers of price predictability: social activity volumes, market share, coin lifetimes, variance, etc. Our analysis has identified several trends and potential explanatory variables however we focused on the three coins with the highest community and developer interest. An expanded analysis that focuses on a larger variety of coins along both axes would provide a more rigorous evaluation.

\bibliographystyle{./aaai}
\bibliography{crypto_bibliography}

\end{document}

%% file: figs/external_plus_activity_over_time_NCOMMENTS_BTC.tex
	\begin{tikzpicture}
	\begin{axis}[ 
	ylabel = \small \# Comments , ylabel style ={yshift=0.05in},
	height = 2in, width = 6.5in, 
	xtick align = center, ytick align = center, 
	title style = {yshift = -.1in,align = center}, 
	xtick pos = left,  
	xmin=0,xmax=36,
	xtick ={0, 1, 2, 3, 4, 5, 6, 7, 8, 9, 10, 11, 12, 13, 14, 15, 16, 17, 18, 19, 20, 21, 22, 23, 24, 25, 26, 27, 28, 29, 30, 31, 32, 33, 34, 35, 36, 37},
	xticklabels = {Jan 2015, , , , , , , , , , , , Jan 2016,  , , ,  , , , , , , , ,Jan 2017,  , , , , , , , , , , ,Jan 2018, }, 
	xtick={0,6,12,18,24,30,36},
	xticklabels = {Jan 2015,Jun 2015,Jan 2016,Jun 2016,Jan 2017,Jun 2017,Jan 2018}, 
	xticklabels = {\textcolor{white}{January 2015},\textcolor{white}{June 2015},\textcolor{white}{January 2016},\textcolor{white}{June 2016},\textcolor{white}{January 2017},\textcolor{white}{June 2017},\textcolor{white}{January 2018}},  
	ytick ={0,200000,400000,600000,800000},
	yticklabels ={\small 0,\small 200K,\small 400K, \small 600K, \small 800K},
	ymin=0,
	ymax=800000,
	scaled y ticks=false,
	xmax=31.5,
	]
	\scriptsize
	\tiny
	
	\draw [-,dashed] (axis cs:36,0000) -- (axis cs:36,300000) -- (axis cs:37,300000)-- (axis cs:37,0000);
	\node[above,align=center] at (axis cs:37,300000) { BTC\\\textcolor{red!75!black}{\textbf{-}\$10K }\\\textcolor{red!75!black}{ (57\%)}}; 
	
	\draw [-,dashed] (axis cs:35,0000) -- (axis cs:35,600000) -- (axis cs:32,600000)-- (axis cs:32,0000);
	\node[above,align=center] at (axis cs:34,600000) { BTC \textcolor{green!50!black}{\textbf{\tiny +}\$13K (275\%)}};
	
	\draw [-,dashed] (axis cs:31,0000) -- (axis cs:31,440000);
	\node[above,align=center] at (axis cs:31,400000) { BTC\\ \textcolor{green!50!black}{\textbf{\tiny +}\$1K } \\ \textcolor{green!50!black}{\textbf(59\%) } }; 
	\draw [-,dashed] (axis cs:31,570000) -- (axis cs:31,690000);
	\node[left,align=center] at (axis cs:31,680000) { BTC splits into BTC and BCH  };

	\draw [-,dashed] (axis cs:30,0) -- (axis cs:30,300000) -- (axis cs:29,300000) -- (axis cs:29,0) ;
	\node[above,align=center] at (axis cs:29.5,300000) {BTC\\ \textcolor{red!75!black}{\textbf{-}\$600} \\  \textcolor{red!75!black}{\textbf(22\%)}}; 
	
	\draw [-,dashed] (axis cs:28,0) -- (axis cs:28,150000);
	\node[above,align=center] at (axis cs:28,150000) {BTC\\ \textcolor{green!50!black}{\textbf{\tiny +}\$1K} \\ \textcolor{green!50!black}{\textbf(61\%)}};  
	
	\draw [-,dashed] (axis cs:27,0) -- (axis cs:27,550000);
	\node[left,align=right] at (axis cs:27,550000) {April 1 - Japan legalizes \\ bitcoin as a method \\ of payment};  
	
	\draw [-,dashed] (axis cs:26,0) -- (axis cs:26,350000);
	\node[left,align=left] at (axis cs:26.5,350000) {SEC denies  applications to \\ operate exchange-traded \\ fund (ETF) for Bitcoin};  
	
	\draw [-,dashed] (axis cs:19,0) -- (axis cs:19,100000);
	\node[above,align=left] at (axis cs:21.5,100000) {Bitfinex (largest bitcoin \\ exchange by volume) hacked};  %

	\draw [-,dashed] (axis cs:17,0) -- (axis cs:17,150000);
	\node[above left,align=right] at (axis cs:17.5,150000) {HashOcean (major bitcoin miner)\\ disappears with millions of dollars \\ worth of Bitcoin};

	\draw [-,dashed] (axis cs:8,0) -- (axis cs:8,600000);
	\draw [-,dashed] (axis cs:7.8,600000) -- (axis cs:8.2,600000);
	\node[right,align=left] at (axis cs:8,600000) {\$30M raised from investors \\ including Visa, Capital One for \\ Bitcoin startup company };   
	\draw [-,dashed] (axis cs:7.8,400000) -- (axis cs:8.2,400000);
	\node[right,align=left] at (axis cs:8,400000) {22 investment banks have joined \\ R3's ``blockchain consortium'' };   
	
	\draw [-,dashed] (axis cs:7,0) -- (axis cs:7,600000);
	\node[left,align=right] at (axis cs:7,600000) {Bitcoin XT created from\\  fork of software that \\ runs  Bitcoin network to \\ increase block size limit};   
	
	\draw [-,dashed] (axis cs:5,0) -- (axis cs:5,200000);
	\node[above left,align=right] at (axis cs:6,200000) {New York passes first \\ state-specific licensing \\ for digital currencies};   
	
	\addplot[no marks, reddit, thick]
	coordinates{
		(0,156382)
		(1,111593)
		(2,110055)
		(3,102787)
		(4,104558)
		(5,102329)
		(6,111071)
		(7,91297)
		(8,61475)
		(9,63467)
		(10,69347)
		(11,76520)
		(12,79049)
		(13,55352)
		(14,53633)
		(15,40310)
		(16,56373)
		(17,67734)
		(18,54189)
		(19,69894)
		(20,36803)
		(21,46988)
		(22,57021)
		(23,70452)
		(24,86271)
		(25,82755)
		(26,132130)
		(27,69076)
		(28,136341)
		(29,117596)
		(30,124201)
		(31,199417)
		(32,167813)
		(33,196488)
		(34,388390)
		(35,583783)
		(36,261390)
		(37,190146)
	};

	\end{axis}
	\end{tikzpicture}
	\vspace{-0.1in}

	\begin{tikzpicture}
	\begin{axis}[ 
	ylabel = \# events, ylabel style={yshift=0.1in},
	height = 1in, width = 6.5in, 
	xtick align = center, ytick align = center, 
	title style = {yshift = -.1in,align = center}, 
	xtick pos = left,  
	xmin=0,xmax=36,
	xtick ={0, 1, 2, 3, 4, 5, 6, 7, 8, 9, 10, 11, 12, 13, 14, 15, 16, 17, 18, 19, 20, 21, 22, 23, 24, 25, 26, 27, 28, 29, 30, 31, 32, 33, 34, 35, 36, 37}, 
	xticklabels = {Jan 2015, , , , , , , , , , , , Jan 2016,  , , ,  , , , , , , , ,Jan 2017,  , , , , , , , , , , ,Jan 2018, }, 
	xtick={0,6,12,18,24,30,36},
	xticklabels = {Jan 2015,Jun 2015,Jan 2016,Jun 2016,Jan 2017,Jun 2017,Jan 2018}, 
	xticklabels = {January 2015,June 2015,January 2016,June 2016,January 2017,June 2017,January 2018},  
	ymin=0,
	ymax=6000,
	scaled y ticks=false, 
	xmax=31.5,
	xticklabels = {\textcolor{white}{January 2015},\textcolor{white}{June 2015},\textcolor{white}{January 2016},\textcolor{white}{June 2016},\textcolor{white}{January 2017},\textcolor{white}{June 2017},\textcolor{white}{January 2018}},  
	ytick = {0,2000,4000,6000}, yticklabels={\small 0, 2K, 4K, 6K},
	]
	\small
	\draw [-,dashed] (axis cs:36,0000) -- (axis cs:36,6000) -- (axis cs:37,6000)-- (axis cs:37,0000); 
	\draw [-,dashed] (axis cs:35,0000) -- (axis cs:35,6000) -- (axis cs:32,6000)-- (axis cs:32,0000); 
	\draw [-,dashed] (axis cs:31,0000) -- (axis cs:31,6000);  
	\draw [-,dashed] (axis cs:30,0) -- (axis cs:30,6000) -- (axis cs:29,6000) -- (axis cs:29,0) ; 
	\draw [-,dashed] (axis cs:28,0) -- (axis cs:28,6000); 
	\draw [-,dashed] (axis cs:27,0) -- (axis cs:27,6000); 
	\draw [-,dashed] (axis cs:26,0) -- (axis cs:26,6000); 
	\draw [-,dashed] (axis cs:19,0) -- (axis cs:19,6000); 
	\draw [-,dashed] (axis cs:17,0) -- (axis cs:17,6000); 	
	\draw [-,dashed] (axis cs:8,0) -- (axis cs:8,6000); 
	\draw [-,dashed] (axis cs:7,0) -- (axis cs:7,6000); 
	\draw [-,dashed] (axis cs:5,0) -- (axis cs:5,6000); 
	
	\addplot[github,no marks, thick] coordinates{
		(0,1842)
		(1,1572)
		(2,1633)
		(3,1683)
		(4,1937)
		(5,2307)
		(6,2327)
		(7,1822)
		(8,2214)
		(9,2816)
		(10,3507)
		(11,2395)
		(12,2757)
		(13,2209)
		(14,2427)
		(15,2454)
		(16,2276)
		(17,2605)
		(18,2008)
		(19,2651)
		(20,2924)
		(21,2508)
		(22,2834)
		(23,2626)
		(24,3490)
		(25,3180)
		(26,3488)
		(27,2698)
		(28,3408)
		(29,4167)
		(30,4683)
		(31,4930)
	}; 
	\end{axis}
	\end{tikzpicture}
	
	\vspace{-0.15in}
	\begin{tikzpicture}
	\begin{axis}[ 
	xlabel = Month,
	ylabel = USD (log), 
	height = 1in, width = 6.5in, 
	xtick align = center, ytick align = center, 
	title style = {yshift = -.1in,align = center}, 
	xtick pos = left,  
	xmin=0,xmax=36,
	xtick ={0, 1, 2, 3, 4, 5, 6, 7, 8, 9, 10, 11, 12, 13, 14, 15, 16, 17, 18, 19, 20, 21, 22, 23, 24, 25, 26, 27, 28, 29, 30, 31, 32, 33, 34, 35, 36, 37}, 
	xticklabels = {Jan 2015, , , , , , , , , , , , Jan 2016,  , , ,  , , , , , , , ,Jan 2017,  , , , , , , , , , , ,Jan 2018, }, 
	xtick={0,6,12,18,24,30,36},
	xticklabels = {Jan 2015,Jun 2015,Jan 2016,Jun 2016,Jan 2017,Jun 2017,Jan 2018}, 
	xticklabels = {January 2015,June 2015,January 2016,June 2016,January 2017,June 2017,January 2018},  
	ymin=100,
	ymax=10000,
	ytick={100,1000,10000}, 
	yticklabels={100,1K,10K},
	ymode=log,
	scaled y ticks=false, 
	xmax=31.5,
	]
	\small
	\draw [-,dashed] (axis cs:36,0000) -- (axis cs:36,100000) -- (axis cs:37,100000)-- (axis cs:37,0000); 
	\draw [-,dashed] (axis cs:35,0000) -- (axis cs:35,100000) -- (axis cs:32,100000)-- (axis cs:32,0000); 
	\draw [-,dashed] (axis cs:31,0000) -- (axis cs:31,100000);  
	\draw [-,dashed] (axis cs:30,0) -- (axis cs:30,100000) -- (axis cs:29,100000) -- (axis cs:29,0) ; 
	\draw [-,dashed] (axis cs:28,0) -- (axis cs:28,100000); 
	\draw [-,dashed] (axis cs:27,0) -- (axis cs:27,100000); 
	\draw [-,dashed] (axis cs:26,0) -- (axis cs:26,100000); 
	\draw [-,dashed] (axis cs:19,0) -- (axis cs:19,100000); 
	\draw [-,dashed] (axis cs:17,0) -- (axis cs:17,100000); 	
	\draw [-,dashed] (axis cs:8,0) -- (axis cs:8,100000); 
	\draw [-,dashed] (axis cs:7,0) -- (axis cs:7,100000); 
	\draw [-,dashed] (axis cs:5,0) -- (axis cs:5,100000); 
	\addplot[blue,no marks, thick] coordinates{
		(0,316.4)
		(1, 242.51)
		(2, 274.8)
		(3, 254.85)
		(4, 236.38)
		(5, 227.44)
		(6, 261.29)
		(7, 282.5)
		(8, 230.59)
		(9, 238.45)
		(10, 365.75)
		(11, 363.43)
		(12, 437.43)
		(13, 376.63)
		(14, 435.91)
		(15, 420.09)
		(16, 453.87)
		(17, 541.55)
		(18, 700.46)
		(19, 613.75)
		(20, 577.6)
		(21, 614.21)
		(22, 735.78)
		(23, 779.02)
		(24, 1031.68)
		(25, 1009.13)
		(26, 1286.98)
		(27, 1109.48)
		(28, 1471.14)
		(29, 2493.97)
		(30, 2555.34)
		(31, 2773.8)
		(32, 4976.52)
		(33, 4477.38)
		(34, 7339.91)
		(35, 11175.23)
		(36, 15306.13)
		(37, 9147.93)
	}; 
	\end{axis}
	\end{tikzpicture}
	
	\small
	\begin{tikzpicture} 
	\begin{axis}[%
	hide axis,
	xmin=10,xmax=50,ymin=0,ymax=0.4,
	legend style={draw=none,legend cell align=left,legend columns = -1, column sep = 3mm}
	]  
	\addlegendimage{reddit, no marks, very thick}; 
	\addlegendentry{Reddit Social Interactions}; 
	\addlegendimage{github, no marks, very thick}; 
	\addlegendentry{GitHub Social Interactions}; 
	\addlegendimage{price, no marks, very thick}; 
	\addlegendentry{Price High}; 
	\end{axis}
	\end{tikzpicture}
	\vspace{-0.1in}
	
	

%% file: figs/lstm_architecture_KDD.tex
 \begin{tikzpicture}[scale=0.45, transform shape]
 
 \draw[->,ultra thick] (0.5,-0.6) -- (0.5,0);
 \node[above] at (0.5,0) {\huge Price Forecast};%
 
 
 
 \draw  (-0.2,-0.8) rectangle (1.2,-1.7);
 \node[right] at (3,-1.4) {\huge Dense Layer (1)};%

 
 \draw (-1.25,-2.4) edge (0.1,-1.7) {};
 \draw (2.25,-2.4) edge (0.9,-1.7) {};
 
 \draw  (1.75,-2.4) edge (0.3,-1.7);
 \draw  (-0.75,-2.4) edge (0.7,-1.7);
 
 \draw  (-1.75,-2.4) rectangle (-0.75,-3.2); 
 \draw  (1.75,-2.4) rectangle (2.75,-3.2); 
 \node[right] at (3,-2.9) {\huge LSTM Layer (800)};
 
 \draw[->,thick] (-0.4,-2.8) -- (0,-2.8);
 \draw[loosely dotted, ultra thick] (0.1,-2.8) edge (0.9,-2.8);
 \draw[->,thick] (0.9,-2.8) -- (1.4,-2.8);
 
 \draw  (-1.75,-3.8) rectangle (-0.75,-4.6); 
 \draw  (1.75,-3.8) rectangle (2.75,-4.6); 
 \node[right] at (3,-4.3) {\huge LSTM Layer (400)};%
 
 \draw (-1.25,-3.8) edge (-1.25,-3.2) {};
 \draw (2.25,-3.8) edge (2.25,-3.2) {};
 
 \draw  (-1,-3.2) edge (2,-3.8);
 \draw  (-1,-3.8) edge (2,-3.2);
 
 \draw[->,thick] (-0.4,-4.2) -- (0,-4.2);
 \draw[loosely dotted, ultra thick] (0.1,-4.2) edge (0.9,-4.2);
 \draw[->,thick] (0.9,-4.2) -- (1.4,-4.2);
 
 \draw[->,thick] (0.5,-5) -- (0.5,-4.5);
 \node[below] at (3.75,-5) {\huge Input};%
 
 \draw  (-1,-5) rectangle (-0.5,-5.5);
 \draw  (-0.5,-5) rectangle (0,-5.5);
 \draw  (0,-5) rectangle (0.5,-5.5);
 \draw  (0.5,-5) rectangle (1,-5.5);
 \draw  (1,-5) rectangle (1.5,-5.5);
 \draw  (1.5,-5) rectangle (2,-5.5);

 \end{tikzpicture}

%% file: figs/performance_by_FW_twoweeks_plot.tex
\pgfplotstableread{
	featureset	FW	bitcoinRMSE	bitcoinMAPE	bitcoinMAPECIwidth	bitcoinMAXAPE	bitcoinRMSPE	bitcoinRMSPECIwidth	ethereumRMSE	ethereumMAPE	ethereumMAPECIwidth	ethereumMAXAPE	ethereumRMSPE	ethereumRMSPECIwidth	moneroRMSE	moneroMAPE	moneroMAPECIwidth	moneroMAXAPE	moneroRMSPE	moneroRMSPECIwidth
	ARIMA	1	600781.9139	3.235296203	0.649848352	23.18638291	0.482539628	0.335206076	7657.375696	5.133862042	1.18126982	48.19208341	0.828904415	0.649889053	1037.552528	5.838659037	1.191248803	42.70894057	0.878408914	0.626107643
	ARIMA	2	910990.1655	5.34805689	0.870554911	23.35487522	0.718357059	0.412599482	11125.80616	8.438963031	1.561864382	48.3517541	1.205217423	0.782939885	1431.542246	8.741200323	1.606543106	41.13328961	1.243960961	0.755913163
	ARIMA	3	1149954.022	6.797538828	1.029775559	23.94071144	0.885391785	0.472179832	13924.54031	10.8535951	1.829435955	50.02587875	1.481149176	0.873953981	1811.107113	11.13881125	1.992072273	49.89165242	1.563702338	0.917126854
	ARIMA	4	1372655.412	8.104172516	1.1802371	29.9033265	1.03901543	0.540810495	16012.42008	12.65636065	2.033422474	49.20947094	1.6902039	0.956976579	2101.111975	13.08307138	2.221286918	49.36205571	1.791429644	0.988132081
	ARIMA	5	1547378.071	9.294569189	1.286636982	30.20644249	1.168901571	0.578499325	18064.73004	14.45322527	2.132354698	50.27828803	1.862524408	0.989218019	2349.144934	14.92907056	2.290603684	50.65393223	1.954800931	1.025700392
	ARIMA	6	1688524.287	10.1546296	1.391828815	30.45955739	1.272446645	0.615380819	19904.42881	16.43130258	2.262260826	51.69637864	2.062325623	1.048034734	2695.529069	16.23114336	2.447857443	61.71836758	2.110244752	1.107897643
	ARIMA	7	1805992.226	10.86934387	1.470350636	34.04645733	1.355578725	0.652344599	21965.72779	18.46846068	2.365481285	54.98149606	2.260338514	1.100987466	3010.650685	17.56457332	2.606191648	60.21855734	2.268623471	1.175624922
	ARIMA	8	1933850.461	11.58991915	1.544130803	34.1915932	1.437683164	0.68811708	23860.52499	20.15572117	2.508687255	58.01962733	2.443905413	1.142927293	3307.244097	18.70674941	2.738430013	66.51813316	2.403215484	1.240830871
	ARIMA	9	2068253.364	12.39448793	1.607211916	36.14900034	1.52323298	0.721719834	25747.94732	21.71989339	2.6775884	57.23945406	2.625556182	1.196207905	3498.487449	19.56160462	2.832125192	66.16423617	2.5021982	1.294391404
	ARIMA	10	2180311.012	13.06926704	1.659522127	35.4263209	1.594969482	0.743627018	27391.59187	23.32939565	2.775906457	62.41673153	2.789507052	1.233966295	3682.541163	20.16716632	2.88245559	67.86487904	2.56687995	1.324743054
	ARIMA	11	2270677.505	13.71515241	1.673421725	34.96109615	1.652569172	0.76023147	28908.92256	24.88821173	2.843101234	62.3848395	2.940674923	1.266282176	3844.734354	20.55140051	2.902452607	67.525207	2.603927347	1.340841197
	ARIMA	12	2372252.491	14.45139721	1.699451677	35.28151271	1.721919421	0.774692993	30613.21411	26.40607855	2.940157272	67.58267634	3.097823128	1.314646701	4015.658675	20.94224006	2.928806079	67.7305898	2.643719488	1.361493297
	ARIMA	13	2485005.201	15.25365753	1.718811534	36.31851009	1.795384014	0.782793501	32190.74757	27.89562137	3.012232111	69.05207149	3.2458537	1.352024892	4026.731134	20.94980627	2.897627619	66.88215518	2.633873005	1.362288329
	ARIMA	14	2585155.5	15.85131765	1.775271019	37.31214603	1.862573601	0.799605102	33672.47221	29.33690145	3.015414754	74.69613891	3.371389873	1.386706063	4062.094271	21.36189162	2.885386329	67.04756103	2.662735696	1.36741759
}{\arimaTable}

\pgfplotstableread{
	featureset	FW	bitcoinRMSE	bitcoinMAPE	bitcoinMAPECIwidth	bitcoinMAXAPE	bitcoinRMSPE	bitcoinRMSPECIwidth	ethereumRMSE	ethereumMAPE	ethereumMAPECIwidth	ethereumMAXAPE	ethereumRMSPE	ethereumRMSPECIwidth	moneroRMSE	moneroMAPE	moneroMAPECIwidth	moneroMAXAPE	moneroRMSPE	moneroRMSPECIwidth
	norm_price_high	1	121.1998932	3.110308306	0.55344754	17.15859348	0.434636896	0.268801462	18.16554083	4.709071211	0.963933696	26.18019525	0.708058379	0.448495147	6.61467926	5.708314562	1.119946159	42.52987539	0.838611905	0.595630347
	norm_price_high	2	183.2740974	5.141682582	0.756470589	22.03138448	0.660728627	0.367723371	33.77436054	9.806753034	1.493789964	37.29635584	1.277955532	0.683924606	9.192867025	8.441526188	1.552526046	41.35397467	1.19911668	0.74013678
	norm_price_high	3	230.9248574	6.325562312	0.917203572	23.01238679	0.808250628	0.428658256	36.18167483	10.35733048	1.659582594	37.1733428	1.37895075	0.734510594	11.79992816	10.76613572	1.92319051	50.23357595	1.507332004	0.889610303
	norm_price_high	4	277.2313642	7.427797504	1.061559222	29.78831588	0.943830594	0.494973824	44.7485237	12.84213863	1.993365835	50.29497814	1.686671629	0.891921497	14.21818067	12.8730893	2.182290218	50.85095898	1.757896659	0.977866341
	norm_price_high	5	305.8242622	8.496853714	1.137198451	27.48364402	1.054089274	0.506531166	50.69231515	14.59478217	2.180367512	53.68080892	1.886952307	0.975343221	16.34494855	14.53569696	2.408828075	53.09873546	1.964398927	1.059718418
	norm_price_high	6	329.4939844	9.363423613	1.157389187	27.60547311	1.131289896	0.529594594	53.88160569	16.00853307	2.268626017	54.90037509	2.027658311	1.024865946	17.55633253	15.77932076	2.400809937	61.92147772	2.055301351	1.102343258
	norm_price_high	7	350.1209179	9.981079625	1.190244232	30.32045548	1.192689712	0.543865972	58.99211869	17.87636361	2.372995636	58.02796102	2.211353279	1.08567004	19.7456559	16.88167164	2.62598326	61.52980904	2.219211316	1.194250096
	norm_price_high	8	371.7325512	10.64247354	1.220995659	31.60228056	1.257467154	0.574258587	63.76416542	19.57027984	2.509365395	60.97648723	2.392643697	1.141644121	21.69102805	18.20032729	2.776908855	68.0701344	2.373368827	1.272089584
	norm_price_high	9	401.7791259	11.42590894	1.293189137	35.20886114	1.344891113	0.622935135	71.32482673	22.26286442	2.762495255	61.51790945	2.693081235	1.235257743	23.66604308	19.47073837	2.968356256	68.72548544	2.538194114	1.34617362
	norm_price_high	10	438.3174882	12.03234567	1.428308539	34.88712853	1.43584234	0.659879345	73.82903091	23.19671844	2.818988479	59.73622518	2.787847459	1.257098674	23.691808	19.1769186	2.94961198	68.58206909	2.509082679	1.342565325
	norm_price_high	11	457.3521994	12.54563934	1.450808722	33.68906433	1.485696274	0.680532836	76.96289319	24.52244452	2.908838938	61.67089982	2.925675985	1.294881369	25.29071498	20.07108571	3.067360798	69.16837682	2.619089273	1.390796587
	norm_price_high	12	488.0664086	13.42632568	1.531119143	35.37685463	1.583693823	0.710864324	79.48249824	25.51532441	2.971853579	61.53860873	3.027854343	1.323213489	24.66390984	20.00015796	2.924556216	67.97288006	2.563928429	1.352756774
	norm_price_high	13	517.5768247	14.28742919	1.575134005	36.68683333	1.669691777	0.725148483	85.55053452	27.66451859	3.085982653	64.69297478	3.24328223	1.380862616	25.22074094	20.00812758	2.955215448	68.08052556	2.575104226	1.378807555
	norm_price_high	14	561.1327053	15.34724427	1.69548495	37.15428599	1.79454367	0.760102884	94.13266364	30.38930835	3.29275628	66.64962429	3.535195477	1.468356111	25.76202173	21.38595535	2.949876118	69.26165317	2.681792978	1.399703512
}{\lstmPriceTable}
\pgfplotstableread{
	featureset	FW	bitcoinRMSE	bitcoinMAPE	bitcoinMAPECIwidth	bitcoinMAXAPE	bitcoinRMSPE	bitcoinRMSPECIwidth	ethereumRMSE	ethereumMAPE	ethereumMAPECIwidth	ethereumMAXAPE	ethereumRMSPE	ethereumRMSPECIwidth	moneroRMSE	moneroMAPE	moneroMAPECIwidth	moneroMAXAPE	moneroRMSPE	moneroRMSPECIwidth
	norm_price_high+R_comments	1	124.0629631	3.258112608	0.551704325	16.66159818	0.444682278	0.268778211	18.44614304	4.756617546	0.974799727	26.52535787	0.715671896	0.454425491	6.561083828	5.936858334	1.10645423	41.8251057	0.84902619	0.585061321
	norm_price_high+R_comments	2	190.5010501	5.375680461	0.789696898	23.78871718	0.690384928	0.382252383	26.82656658	7.674026088	1.336426234	33.64073868	1.061290306	0.606683631	8.948387127	8.902563771	1.498672411	39.80435199	1.211775194	0.72628614
	norm_price_high+R_comments	3	230.4336707	6.366539106	0.919617832	23.91613966	0.812283378	0.431317033	33.21997203	9.858878119	1.532205972	35.41094753	1.295530973	0.690428443	11.52642124	10.95691389	1.895080279	49.35729825	1.510362711	0.885239471
	norm_price_high+R_comments	4	271.1345017	7.357331684	1.057120307	28.88944891	0.936785892	0.48963603	39.57068889	11.41563875	1.804936412	47.75441835	1.511112617	0.825187989	13.49331939	12.79268865	2.120423463	49.14214156	1.729007029	0.960429014
	norm_price_high+R_comments	5	312.7369043	8.599589731	1.144472289	28.36649565	1.064735839	0.516837684	47.40555922	13.75249471	2.053841756	52.55239346	1.77781231	0.939988685	15.53528967	14.37383137	2.28515274	51.49427943	1.907192483	1.025824843
	norm_price_high+R_comments	6	338.1370029	9.328134303	1.211293794	28.30995064	1.145269132	0.539494376	55.06950402	16.37217611	2.311636557	55.45125796	2.070851747	1.038712816	17.52889025	15.78499167	2.397781889	61.88453032	2.05467305	1.101515934
	norm_price_high+R_comments	7	363.9634464	10.0033434	1.256788452	27.86316634	1.214889494	0.55909943	57.66103254	17.67516184	2.340153672	58.08051593	2.184485275	1.081302158	19.71466171	16.87400819	2.622039944	61.47843329	2.217224586	1.19307611
	norm_price_high+R_comments	8	365.3340434	10.57727504	1.193123549	31.99429853	1.243841502	0.56906678	67.40786264	20.6656454	2.628753068	62.29590646	2.519932355	1.180398777	21.54223572	18.21336148	2.770095294	67.81387345	2.371972404	1.267424578
	norm_price_high+R_comments	9	402.1423543	11.41384698	1.295702964	35.21586038	1.344594918	0.623100468	67.31546489	21.10145275	2.620935857	59.92235275	2.553377089	1.19062663	23.03960312	18.80921115	2.91347477	67.73385464	2.46821053	1.323449388
	norm_price_high+R_comments	10	440.120189	12.081149	1.43773005	35.04143661	1.442753141	0.662324128	71.70164174	22.77137723	2.756534118	59.28668049	2.733457825	1.24279487	23.5569142	19.15619935	2.927885533	68.33210642	2.499825714	1.337887342
	norm_price_high+R_comments	11	451.1833705	12.64994718	1.425391391	33.73759388	1.48713701	0.677071559	76.72701709	24.46842984	2.901525786	61.60560689	2.918960696	1.292742194	25.20363672	19.98581075	3.054717373	69.01684556	2.608098677	1.38670759
	norm_price_high+R_comments	12	485.5414187	13.39818119	1.517433682	35.2473024	1.577333956	0.708307912	79.2888995	25.47555832	2.96787888	61.49610513	3.023329429	1.32188779	24.67986154	19.96764736	2.924991797	68.01723872	2.561542877	1.353540773
	norm_price_high+R_comments	13	515.2560851	14.25568639	1.564645046	36.576184	1.664001344	0.722686132	85.08798966	27.51997135	3.079614101	64.55039702	3.22913248	1.376433469	25.88062241	20.96367149	3.031296116	69.22024518	2.675764511	1.410700265
	norm_price_high+R_comments	14	558.612217	15.29430197	1.687176482	37.04890429	1.787654255	0.757399153	93.83705574	30.30957293	3.286059261	66.58759632	3.526464385	1.466306824	25.73793427	21.44260496	2.951280989	69.19176795	2.686776957	1.398837601
}{\lstmPriceRlangTable}
\pgfplotstableread{
	featureset	FW	bitcoinRMSE	bitcoinMAPE	bitcoinMAPECIwidth	bitcoinMAXAPE	bitcoinRMSPE	bitcoinRMSPECIwidth	ethereumRMSE	ethereumMAPE	ethereumMAPECIwidth	ethereumMAXAPE	ethereumRMSPE	ethereumRMSPECIwidth	moneroRMSE	moneroMAPE	moneroMAPECIwidth	moneroMAXAPE	moneroRMSPE	moneroRMSPECIwidth
	norm_price_high+GH_pop	1	136.6245912	3.685238847	0.553070687	15.75337451	0.477339952	0.272448859	18.37802903	4.748114732	0.970214428	26.59142048	0.713228199	0.453710027	6.759177548	5.734491846	1.142779023	42.96831514	0.849594932	0.608044689
	norm_price_high+GH_pop	2	186.660933	5.251315417	0.762495071	22.52547497	0.671349425	0.370675015	31.86600661	9.049015767	1.475732562	36.25951373	1.21414907	0.66430811	9.13845475	8.424549058	1.54939752	40.99708518	1.196702678	0.743432364
	norm_price_high+GH_pop	3	234.7350206	6.54204659	0.938240233	24.93311634	0.83238845	0.440455408	39.36858661	11.29173473	1.756656201	38.68673024	1.484447273	0.77075165	11.88026344	10.74719297	1.940295598	50.16258958	1.512567377	0.890079535
	norm_price_high+GH_pop	4	272.0361983	7.425718498	1.045775357	29.83557385	0.938348897	0.491179587	43.6304191	12.52480596	1.952779226	49.78940567	1.648080742	0.879434106	14.18237487	12.94653673	2.183071619	50.89461154	1.763573343	0.977145086
	norm_price_high+GH_pop	5	302.500566	8.533416292	1.116048153	27.15702281	1.050233608	0.500782728	50.05362896	14.61362832	2.170097741	54.27292999	1.884847384	0.980401255	15.6395532	14.40142024	2.320555553	52.3119046	1.922078647	1.035264132
	norm_price_high+GH_pop	6	331.7545886	9.41549219	1.183402646	27.92285957	1.14364225	0.532907112	53.72224864	15.96095877	2.256268901	54.65754301	2.019743513	1.021879796	17.55816466	15.65645507	2.396715354	61.71877028	2.044438526	1.100233746
	norm_price_high+GH_pop	7	350.9326172	9.962301526	1.195023485	29.30758255	1.192557772	0.542506465	58.00241087	17.63240887	2.343411026	57.76951131	2.182079633	1.076045432	19.81090997	17.21756544	2.642692556	61.19578427	2.250761358	1.201606137
	norm_price_high+GH_pop	8	364.185087	10.65248866	1.181721058	32.01627156	1.246980775	0.567766225	63.05226301	19.38147593	2.476481695	60.93789895	2.366826313	1.131690629	21.71500679	18.2004394	2.788962281	68.26588894	2.377626461	1.274175266
	norm_price_high+GH_pop	9	396.5821575	11.4344961	1.257472774	35.04893611	1.335395983	0.618916189	67.74448584	21.21422509	2.641009634	60.19459609	2.568898378	1.196371887	23.15640981	19.25815911	2.949159978	67.93824045	2.515137859	1.331617632
	norm_price_high+GH_pop	10	427.7204715	12.17143335	1.35783407	34.78360116	1.426963224	0.650813327	75.84797953	23.9295567	2.879276555	60.96663009	2.867198219	1.28372848	24.66521418	20.31137457	3.087215634	69.80024517	2.644510056	1.385136806
	norm_price_high+GH_pop	11	443.0144988	12.85640142	1.359259571	35.57421804	1.486210685	0.675061144	78.96905469	24.97069796	2.992721451	61.86360117	2.988376645	1.314421148	25.38313499	20.34583616	3.107969858	69.44876003	2.654455509	1.400961527
	norm_price_high+GH_pop	12	459.9080706	13.14406444	1.394314414	34.82369651	1.520743667	0.680566544	79.79875492	25.56477251	2.980738826	61.52456796	3.034645796	1.32494598	25.84997572	21.39291853	3.068204928	70.468263	2.721997778	1.413579844
	norm_price_high+GH_pop	13	533.4611907	14.71348453	1.680669192	36.51842975	1.736324491	0.746861385	87.35831914	28.29879267	3.121749315	65.66873055	3.307665449	1.401674501	25.6428671	20.80282732	3.00498205	68.7612106	2.654193093	1.401154138
	norm_price_high+GH_pop	14	505.8654286	14.46190572	1.486508099	34.2095393	1.660235869	0.694252287	98.25568774	31.72258717	3.370479773	67.46416783	3.671725548	1.498819335	26.64963305	23.12382755	3.0762826	70.21692351	2.862647592	1.440811635
}{\lstmPriceGHpopTable}
\pgfplotstableread{
featureset	FW	bitcoinRMSE	bitcoinMAPE	bitcoinMAPECIwidth	bitcoinMAXAPE	bitcoinRMSPE	bitcoinRMSPECIwidth	ethereumRMSE	ethereumMAPE	ethereumMAPECIwidth	ethereumMAXAPE	ethereumRMSPE	ethereumRMSPECIwidth	moneroRMSE	moneroMAPE	moneroMAPECIwidth	moneroMAXAPE	moneroRMSPE	moneroRMSPECIwidth
norm_price_high+GH_pop+R_comments	1	124.3676771	3.191732545	0.562339625	16.77322088	0.443876244	0.267870006	18.25259452	4.731321043	0.965127795	26.35288942	0.710028362	0.450259394	6.632418279	5.710493435	1.118804182	42.69580016	0.838301515	0.597137355
norm_price_high+GH_pop+R_comments	2	191.4482934	5.404759938	0.780425521	23.23007955	0.689482329	0.377553217	28.54662517	7.86521852	1.429774871	35.00881454	1.110741867	0.63271362	9.259209961	8.324711079	1.564118081	41.73195114	1.195477542	0.747650508
norm_price_high+GH_pop+R_comments	3	231.3535456	6.358668663	0.922537724	23.12933555	0.812663093	0.429521357	33.10346467	9.84272622	1.5251114	35.31006599	1.291778441	0.686685285	13.0823289	11.76751634	2.01943018	52.58200842	1.616128854	0.927646344
norm_price_high+GH_pop+R_comments	4	274.6879351	7.481540552	1.051432045	30.10104923	0.944663922	0.494660796	41.19389001	11.84191756	1.877870773	49.0496033	1.569531121	0.852901193	14.08560632	12.71059753	2.171612497	50.57654927	1.742021868	0.974995947
norm_price_high+GH_pop+R_comments	5	315.9249841	8.742824018	1.153128043	29.00840584	1.079113972	0.524156783	49.90027834	14.33357619	2.146740851	53.22712777	1.855058645	0.964058166	15.96620993	14.32235529	2.324394514	51.91433447	1.917561366	1.037057896
norm_price_high+GH_pop+R_comments	6	331.3995397	9.263132286	1.186367178	27.92459474	1.132065727	0.52932278	54.23421054	16.116559	2.287198486	55.2052664	2.042440273	1.029058968	18.53790267	16.01037869	2.541435717	63.52930608	2.122933994	1.145445618
norm_price_high+GH_pop+R_comments	7	356.5660292	9.991387746	1.230597789	28.01794503	1.205803885	0.549725091	57.52122702	17.59785634	2.347890289	58.22838033	2.180738799	1.081825492	20.95382381	18.00877072	2.770787561	64.15140883	2.356544113	1.248087821
norm_price_high+GH_pop+R_comments	8	373.4735772	10.51968886	1.229397412	31.85122982	1.249573808	0.567353723	68.52172999	21.09273037	2.650199701	63.07755752	2.56173188	1.19316753	21.57138932	17.99815183	2.770478516	67.93468549	2.35562334	1.267322135
norm_price_high+GH_pop+R_comments	9	415.9410064	11.5510177	1.369158478	36.0900154	1.377802136	0.636768325	68.17596885	21.32901501	2.657380262	60.20460151	2.583442331	1.201399245	23.24286159	19.35764791	2.956995546	68.07482458	2.525522173	1.334982169
norm_price_high+GH_pop+R_comments	10	424.7992973	12.24143895	1.336371845	34.80470421	1.426854892	0.65076805	76.35389484	24.0477805	2.890325418	61.02955053	2.880404197	1.286586139	24.73567543	20.30461261	3.074671158	70.1731338	2.639588529	1.385293449
norm_price_high+GH_pop+R_comments	11	427.7602446	12.45923098	1.278914759	35.92154711	1.429858268	0.652228099	77.232572	24.53317994	2.936246184	61.51031183	2.934799976	1.298502214	26.63085773	22.2010381	3.225767611	71.29575034	2.839008835	1.450086445
norm_price_high+GH_pop+R_comments	12	523.5723553	14.17211221	1.746537571	37.5589876	1.710664262	0.754193469	79.20372367	25.36102751	2.974878642	61.14672758	3.015760994	1.318723153	26.25145097	21.53126387	3.102141646	70.65295612	2.744388058	1.428449621
norm_price_high+GH_pop+R_comments	13	514.4762955	14.38892519	1.57862044	36.13744429	1.679370288	0.724837427	86.98268738	28.15060336	3.113741134	65.39737224	3.292715139	1.39702566	26.92650571	22.70690377	3.169019374	70.85567102	2.859711181	1.455040727
norm_price_high+GH_pop+R_comments	14	504.0557393	14.44608625	1.476191286	33.68006833	1.656083526	0.690312498	95.93070289	30.90720601	3.327436482	66.83207158	3.589449176	1.479812209	26.14033613	22.12479306	2.99615656	69.75433974	2.756137238	1.417654589
}{\lstmPriceGHpopRlangTable}


\begin{tikzpicture}
\begin{axis}[ 
title={\small Bitcoin}, title style={yshift=-0.1in},
ylabel = MaxAPE , ylabel style = {align=center},
height = 1.25 in, width = 1.4in,
xtick align = center,ytick align = center,
xtick pos= left,ytick pos= left,
xmin = 0, xmax = 15, 
ymin = 10, ymax = 40, 
xtick={0,5,10,15},
every tick label/.append style={font=\tiny}
]
\addplot[arima, thick, densely dotted, no marks, mark=o, mark size=0.8] table [x =FW, y = bitcoinMAXAPE ] {\arimaTable}; 
\addplot[price, no marks,  mark=o, mark size=0.8] table [x =FW, y = bitcoinMAXAPE ] {\lstmPriceTable}; 
\addplot[reddit, no marks,  mark=o, mark size=0.8] table [x =FW, y = bitcoinMAXAPE ] {\lstmPriceRlangTable};  
\end{axis}
\end{tikzpicture}
\hspace{-0.15in}
\begin{tikzpicture}
\begin{axis}[ 
title={\small Ethereum},title style={yshift=-0.1in},
height = 1.25 in, width = 1.4in,
xtick align = center,ytick align = center,
xtick pos= left,ytick pos= left,
xmin = 0, xmax = 15, 
ymin=20,ymax=80, 
xtick={0,5,10,15},
every tick label/.append style={font=\tiny}
]

\addplot[arima,  thick,  densely dotted, no marks, mark=o, mark size=0.8] table [x =FW, y = ethereumMAXAPE ] {\arimaTable}; 
\addplot[price, no marks,  mark=o, mark size=0.8] table [x =FW, y = ethereumMAXAPE ] {\lstmPriceTable}; 
\addplot[reddit, no marks,  mark=o, mark size=0.8] table [x =FW, y = ethereumMAXAPE ] {\lstmPriceRlangTable};

\end{axis}
\end{tikzpicture}
\hspace{-0.15in}
\begin{tikzpicture}
\begin{axis}[ 
title={\small Monero},title style={yshift=-0.1in},
height = 1.25 in, width = 1.4in,
xtick align = center,ytick align = center,
xtick pos= left,ytick pos= left,
xmin = 0, xmax = 15, 
xtick={0,5,10,15},
every tick label/.append style={font=\tiny}
]
\addplot[arima, thick,  densely dotted, no marks, mark=o, mark size=0.8] table [x =FW, y = moneroMAXAPE ] {\arimaTable}; 
\addplot[price, no marks,  mark=o, mark size=0.8] table [x =FW, y = moneroMAXAPE ] {\lstmPriceTable}; 
\addplot[reddit, no marks,  mark=o, mark size=0.8] table [x =FW, y = moneroMAXAPE ] {\lstmPriceRlangTable};

\end{axis}
\end{tikzpicture}

\begin{tikzpicture}
\begin{axis}[ 
ylabel = MAPE , ylabel style = {align=center},
height = 1.25 in, width = 1.4in,
xtick align = center,ytick align = center,
xtick pos= left,ytick pos= left,
xmin = 0, xmax = 15, 
ymin=0,ymax=35, 
ytick={0,10,20,30},yticklabels={\tiny 0, \tiny 10,\tiny 20,\tiny 30},
xtick={0,5,10,15},xticklabels={\tiny 0,\tiny 5, \tiny 10, \tiny 15},
] 

\addplot[arima, only marks, mark=diamond*, mark size=0.5, error bars/.cd, y explicit, y dir=both,] table [x expr=\thisrow{FW}-0.4, y = bitcoinMAPE,y error =bitcoinMAPECIwidth ] {\arimaTable}; 
\addplot[price, only marks, mark=square*, mark size=0.5, error bars/.cd, y explicit, y dir=both,] table [x =FW, y = bitcoinMAPE,y error =bitcoinMAPECIwidth ] {\lstmPriceTable}; 
\addplot[reddit, only marks, mark=*, mark size=0.5, error bars/.cd, y explicit, y dir=both,] table [x expr=\thisrow{FW}+0.4, y = bitcoinMAPE,y error =bitcoinMAPECIwidth ] {\lstmPriceRlangTable};

\end{axis}
\end{tikzpicture}
\hspace{-0.1in}
\begin{tikzpicture}
\begin{axis}[ 
height = 1.25 in, width = 1.45in,
xtick align = center,ytick align = center,
xtick pos= left,ytick pos= left,
xmin = 0, xmax = 15, 
ymin=0,ymax=35, 
ytick={0,10,20,30},yticklabels={,,,},
xtick={0,5,10,15},xticklabels={\tiny 0,\tiny 5, \tiny 10, \tiny 15},
]

\addplot[arima, only marks, mark=diamond*, mark size=0.5, error bars/.cd, y explicit, y dir=both,] table [x expr=\thisrow{FW}-0.4, y = ethereumMAPE,y error =ethereumMAPECIwidth ] {\arimaTable}; 
\addplot[price, only marks, mark=square*, mark size=0.5, error bars/.cd, y explicit, y dir=both,] table [x =FW, y = ethereumMAPE,y error =ethereumMAPECIwidth ] {\lstmPriceTable}; 
\addplot[reddit, only marks, mark=*, mark size=0.5, error bars/.cd, y explicit, y dir=both,] table [x expr=\thisrow{FW}+0.4, y = ethereumMAPE,y error =ethereumMAPECIwidth ] {\lstmPriceRlangTable};

\end{axis}
\end{tikzpicture}
\hspace{-0.1in}
\begin{tikzpicture}
\begin{axis}[ 
height = 1.25 in, width = 1.45in,
xtick align = center,ytick align = center,
xtick pos= left,ytick pos= left,
xmin = 0, xmax = 15, 
ymin=0,ymax=35, 
ytick={0,10,20,30},yticklabels={,,,},
xtick={0,5,10,15},xticklabels={\tiny 0,\tiny 5, \tiny 10, \tiny 15},
]

\addplot[arima, only marks,  mark=diamond*, mark size=0.5, error bars/.cd, y explicit, y dir=both,] table [x =FW, y = moneroMAPE,y error =moneroMAPECIwidth  ] {\arimaTable}; 
\addplot[price,only marks,  mark=square*, mark size=0.5, error bars/.cd, y explicit, y dir=both,] table [x expr=\thisrow{FW}-0.4, y = moneroMAPE,y error =moneroMAPECIwidth ] {\lstmPriceTable}; 
\addplot[reddit,only marks,  mark=*, mark size=0.5, error bars/.cd, y explicit, y dir=both,] table [x expr=\thisrow{FW}+0.4, y = moneroMAPE,y error =moneroMAPECIwidth ] {\lstmPriceRlangTable};

\end{axis}
\end{tikzpicture}
\small Forecasting Window (days)

\begin{tikzpicture} 
\begin{axis}[%
hide axis,
xmin=10,xmax=50,ymin=0,ymax=0.4,
legend style={draw=none,legend cell align=left,legend columns = -1, column sep = 1mm}
] 
\addlegendimage{arima, thick,  densely dotted, no marks, mark size=0.8}; 
\addlegendentry{\small ARIMA~($\$$)};
\addlegendimage{price, no marks}; 
\addlegendentry{\small LSTM~($\$$)};
\addlegendimage{reddit, no marks}; 
\addlegendentry{\small LSTM~($\$+R_{Lang}$)}; 
\end{axis}
\end{tikzpicture}

\begin{tikzpicture} 
\begin{axis}[%
hide axis,
xmin=10,xmax=50,ymin=0,ymax=0.4,
legend style={draw=none,legend cell align=left,legend columns = -1, column sep = 2mm}
]  
\addlegendimage{arima, only marks, mark = diamond*, mark size=2}; 
\addlegendentry{\small ARIMA~($\$$)};
\addlegendimage{price, only marks, mark = square*, mark size=2}; 
\addlegendentry{\small LSTM~($\$$)};
\addlegendimage{reddit, only marks, mark = *, mark size=2}; 
\addlegendentry{\small LSTM~($\$+R_{Lang}$)}; 
\end{axis}
\end{tikzpicture}

%% file: figs/price_and_reddit_language_predictions.tex
\pgfplotstableread{
date	yTrue	coin
2017-05-04	1588.11	bitcoin
2017-05-05	1560.42	bitcoin
2017-05-06	1572.89	bitcoin
2017-05-07	1667.67	bitcoin
2017-05-08	1757.39	bitcoin
2017-05-09	1766.18	bitcoin
2017-05-10	1864.76	bitcoin
2017-05-11	1822.51	bitcoin
2017-05-12	1770.5	bitcoin
2017-05-13	1802.75	bitcoin
2017-05-14	1776.65	bitcoin
2017-05-15	1752.55	bitcoin
2017-05-16	1842.83	bitcoin
2017-05-17	1980.49	bitcoin
2017-05-18	1969.7	bitcoin
2017-05-19	2048.45	bitcoin
2017-05-20	2094.94	bitcoin
2017-05-21	2264.76	bitcoin
2017-05-22	2286.3	bitcoin
2017-05-23	2496.98	bitcoin
2017-05-24	2781.76	bitcoin
2017-05-25	2616.52	bitcoin
2017-05-26	2322.37	bitcoin
2017-05-27	2300.52	bitcoin
2017-05-28	2337.35	bitcoin
2017-05-29	2329.26	bitcoin
2017-05-30	2330.62	bitcoin
2017-05-31	2460.84	bitcoin
2017-06-01	2493.97	bitcoin
2017-06-02	2582.77	bitcoin
2017-06-03	2559.79	bitcoin
2017-06-04	2705.38	bitcoin
2017-06-05	2931.24	bitcoin
2017-06-06	2880.95	bitcoin
2017-06-07	2808.41	bitcoin
2017-06-08	2852.07	bitcoin
2017-06-09	2914.19	bitcoin
2017-06-10	2977.86	bitcoin
2017-06-11	2985.06	bitcoin
2017-06-12	2784.77	bitcoin
2017-06-13	2803.72	bitcoin
2017-06-14	2521.6	bitcoin
2017-06-15	2536.42	bitcoin
2017-06-16	2690.71	bitcoin
2017-06-17	2676.04	bitcoin
2017-06-18	2617.75	bitcoin
2017-06-19	2800.48	bitcoin
2017-06-20	2804.41	bitcoin
2017-06-21	2757.32	bitcoin
2017-06-22	2759.66	bitcoin
2017-06-23	2741.56	bitcoin
2017-06-24	2660.66	bitcoin
2017-06-25	2584.75	bitcoin
2017-06-26	2585.06	bitcoin
2017-06-27	2616.95	bitcoin
2017-06-28	2605.86	bitcoin
2017-06-29	2576.28	bitcoin
2017-06-30	2529.62	bitcoin
2017-07-01	2555.34	bitcoin
2017-07-02	2617.48	bitcoin
2017-07-03	2658.73	bitcoin
2017-07-04	2642.74	bitcoin
2017-07-05	2634.84	bitcoin
2017-07-06	2617.48	bitcoin
2017-07-07	2568.73	bitcoin
2017-07-08	2576.73	bitcoin
2017-07-09	2530.34	bitcoin
2017-07-10	2412.75	bitcoin
2017-07-11	2424.82	bitcoin
2017-07-12	2436.66	bitcoin
2017-07-13	2370.53	bitcoin
2017-07-14	2237.13	bitcoin
2017-07-15	2044.41	bitcoin
2017-07-16	2233.83	bitcoin
2017-07-17	2400.74	bitcoin
2017-07-18	2412.38	bitcoin
2017-07-19	2932.81	bitcoin
2017-07-20	2873.96	bitcoin
2017-07-21	2876.71	bitcoin
2017-07-22	2856.67	bitcoin
2017-07-23	2798.89	bitcoin
2017-07-24	2779.08	bitcoin
2017-07-25	2631.73	bitcoin
2017-07-26	2712.92	bitcoin
2017-07-27	2843.78	bitcoin
2017-07-28	2812.13	bitcoin
2017-07-29	2773.06	bitcoin
2017-07-30	2916.3	bitcoin
2017-07-31	2946.02	bitcoin
2017-08-01	2773.8	bitcoin
2017-08-02	2822.88	bitcoin
2017-08-03	2892.67	bitcoin
2017-08-04	3344.01	bitcoin
2017-08-05	3295.07	bitcoin
2017-08-06	3425.13	bitcoin
2017-08-07	3494.87	bitcoin
2017-08-08	3437.15	bitcoin
2017-08-09	3453.84	bitcoin
2017-08-10	3706.48	bitcoin
2017-08-11	3967.26	bitcoin
2017-08-12	4189.42	bitcoin
2017-08-13	4336.71	bitcoin
2017-08-14	4436.48	bitcoin
2017-08-15	4398.06	bitcoin
2017-08-16	4487.5	bitcoin
2017-08-17	4362.73	bitcoin
2017-08-18	4189.68	bitcoin
2017-08-19	4182.29	bitcoin
2017-08-20	4097.25	bitcoin
2017-08-21	4142.68	bitcoin
2017-08-22	4255.62	bitcoin
2017-08-23	4364.11	bitcoin
2017-08-24	4461.71	bitcoin
2017-08-25	4379.28	bitcoin
2017-08-26	4408.18	bitcoin
2017-08-27	4403.13	bitcoin
2017-08-28	4647.83	bitcoin
2017-08-29	4644.06	bitcoin
2017-08-30	4765.07	bitcoin

}{\bitcoinTrue}

\pgfplotstableread{
coin	date	oneDay	twoDay	threeDay
bitcoin	2017-05-04	1622.44226074	1537.5859375	1493.58764648
bitcoin	2017-05-05	1600.27624512	1654.87670898	1519.5402832
bitcoin	2017-05-06	1572.03723145	1632.39904785	1636.24743652
bitcoin	2017-05-07	1584.75366211	1603.76464844	1613.88098145
bitcoin	2017-05-08	1681.45532227	1616.65881348	1585.38842773
bitcoin	2017-05-09	1773.07519531	1714.72351074	1598.21875
bitcoin	2017-05-10	1782.05493164	1807.65014648	1695.80065918
bitcoin	2017-05-11	1882.81884766	1816.75939941	1788.27636719
bitcoin	2017-05-12	1839.62182617	1918.97998047	1797.34143066
bitcoin	2017-05-13	1786.46899414	1875.15612793	1899.07373047
bitcoin	2017-05-14	1819.42456055	1821.23657227	1855.45837402
bitcoin	2017-05-15	1792.75292969	1854.66699219	1801.79711914
bitcoin	2017-05-16	1768.13061523	1827.61096191	1835.06726074
bitcoin	2017-05-17	1860.3951415999998	1802.63464355	1808.14086914
bitcoin	2017-05-18	2001.22766113	1896.23046875	1783.28491211
bitcoin	2017-05-19	1990.18261719	2039.12438965	1876.43249512
bitcoin	2017-05-20	2070.81738281	2027.91662598	2018.65332031
bitcoin	2017-05-21	2118.44604492	2109.74633789	2007.49780273
bitcoin	2017-05-22	2292.5847168	2158.0859375	2088.94824219
bitcoin	2017-05-23	2314.69018555	2334.85961914	2137.06591797
bitcoin	2017-05-24	2531.10180664	2357.3034668	2313.04248047
bitcoin	2017-05-25	2824.18408203	2577.07299805	2335.38647461
bitcoin	2017-05-26	2654.05151367	2874.83203125	2554.19921875
bitcoin	2017-05-27	2351.71557617	2701.96655273	2850.71557617
bitcoin	2017-05-28	2329.28515625	2394.8972168	2678.56420898
bitcoin	2017-05-29	2367.09594727	2372.12255859	2372.81469727
bitcoin	2017-05-30	2358.78955078	2410.51416016	2350.14038086
bitcoin	2017-05-31	2360.18603516	2402.08007812	2388.36279297
bitcoin	2017-06-01	2493.953125	2403.49804688	2379.96582031
bitcoin	2017-06-02	2528.00756836	2539.34179688	2381.37719727
bitcoin	2017-06-03	2619.32763672	2573.92993164	2516.62963867
bitcoin	2017-06-04	2595.68994141	2666.69067383	2551.06933594
bitcoin	2017-06-05	2745.51782227	2642.6784668	2643.43652344
bitcoin	2017-06-06	2978.26220703	2794.89550781	2619.52563477
bitcoin	2017-06-07	2926.4074707	3031.42797852	2771.10668945
bitcoin	2017-06-08	2851.6418457	2978.72119141	3006.68261719
bitcoin	2017-06-09	2896.63720703	2902.73632812	2954.18579102
bitcoin	2017-06-10	2960.67919922	2948.46411133	2878.50610352
bitcoin	2017-06-11	3026.34814453	3013.5559082	2924.04980469
bitcoin	2017-06-12	3033.77563477	3080.30761719	2988.88183594
bitcoin	2017-06-13	2827.28515625	3087.85864258	3055.36987305
bitcoin	2017-06-14	2846.80932617	2877.98364258	3062.89086914
bitcoin	2017-06-15	2556.41503906	2897.82446289	2853.85449219
bitcoin	2017-06-16	2571.65454102	2602.78393555	2873.61474609
bitcoin	2017-06-17	2730.41308594	2618.26367188	2579.80053711
bitcoin	2017-06-18	2715.31054688	2779.54858398	2595.21435547
bitcoin	2017-06-19	2655.31762695	2764.20336914	2755.82324219
bitcoin	2017-06-20	2843.47143555	2703.25268555	2740.54174805
bitcoin	2017-06-21	2847.52075195	2894.43212891	2679.84448242
bitcoin	2017-06-22	2799.0078125	2898.54760742	2870.23632812
bitcoin	2017-06-23	2801.41821289	2849.24804688	2874.33496094
bitcoin	2017-06-24	2782.77539062	2851.69750977	2825.23608398
bitcoin	2017-06-25	2699.47875977	2832.75341797	2827.67529297
bitcoin	2017-06-26	2621.36450195	2748.1184082	2808.80908203
bitcoin	2017-06-27	2621.68359375	2668.76000977	2724.52319336
bitcoin	2017-06-28	2654.49414062	2669.08422852	2645.49682617
bitcoin	2017-06-29	2643.08349609	2702.41625977	2645.81958008
bitcoin	2017-06-30	2612.65185547	2690.82348633	2679.01171875
bitcoin	2017-07-01	2564.66235352	2659.90869141	2667.4675293
bitcoin	2017-07-02	2591.11254883	2611.16088867	2636.68310547
bitcoin	2017-07-03	2655.03955078	2638.02929688	2588.14135742
bitcoin	2017-07-04	2697.49194336	2702.9699707	2614.89575195
bitcoin	2017-07-05	2681.03466797	2746.09985352	2679.56298828
bitcoin	2017-07-06	2672.90405273	2729.37915039	2722.51293945
bitcoin	2017-07-07	2655.03955078	2721.11914062	2705.86181641
bitcoin	2017-07-08	2604.88500977	2702.9699707	2697.63671875
bitcoin	2017-07-09	2613.11450195	2652.01928711	2679.56298828
bitcoin	2017-07-10	2565.40209961	2660.37915039	2628.8269043
bitcoin	2017-07-11	2444.53710938	2611.9128418	2637.15185547
bitcoin	2017-07-12	2456.93847656	2489.15478516	2588.890625
bitcoin	2017-07-13	2469.10424805	2501.74926758	2466.65991211
bitcoin	2017-07-14	2401.16821289	2514.10449219	2479.19970703
bitcoin	2017-07-15	2264.23535156	2445.11328125	2491.50170898
bitcoin	2017-07-16	2066.67944336	2306.07739258	2422.81005859
bitcoin	2017-07-17	2260.84960938	2105.54663086	2284.38891602
bitcoin	2017-07-18	2432.19873047	2302.64013672	2084.76782227
bitcoin	2017-07-19	2444.15698242	2476.62524414	2280.96679688
bitcoin	2017-07-20	2979.88134766	2488.76904297	2454.18457031
bitcoin	2017-07-21	2919.20141602	3033.07348633	2466.27587891
bitcoin	2017-07-22	2922.03637695	2971.39746094	3008.32177734
bitcoin	2017-07-23	2901.37841797	2974.27856445	2946.89160156
bitcoin	2017-07-24	2841.83276367	2953.28295898	2949.76123047
bitcoin	2017-07-25	2821.42358398	2892.76708984	2928.84936523
bitcoin	2017-07-26	2669.70361328	2872.02661133	2868.57788086
bitcoin	2017-07-27	2753.28125	2717.86767578	2847.92138672
bitcoin	2017-07-28	2888.09228516	2802.78393555	2694.39868164
bitcoin	2017-07-29	2855.4753418	2939.77978516	2778.96264648
bitcoin	2017-07-30	2815.22167969	2906.63134766	2915.40063477
bitcoin	2017-07-31	2962.85522461	2865.72436523	2882.38598633
bitcoin	2017-08-01	2993.50463867	3015.76733398	2841.64477539
bitcoin	2017-08-02	2815.98388672	3046.92211914	2991.08447266
bitcoin	2017-08-03	2866.55297852	2866.49902344	3022.11547852
bitcoin	2017-08-04	2938.49023438	2917.88916016	2842.41650391
bitcoin	2017-08-05	3404.51147461	2991.00244141	2893.59838867
bitcoin	2017-08-06	3353.91650391	3464.85083008	2966.41796875
bitcoin	2017-08-07	3488.40625	3413.38891602	3438.4543456999995
bitcoin	2017-08-08	3560.5625	3550.19262695	3387.18139648
bitcoin	2017-08-09	3500.84057617	3623.6027831999995	3523.48657227
bitcoin	2017-08-10	3518.10766602	3562.8425293000005	3596.63378906
bitcoin	2017-08-11	3779.6706543000005	3580.40917969	3536.09057617
bitcoin	2017-08-12	4050.00756836	3846.56982422	3553.59399414
bitcoin	2017-08-13	4280.55761719	4121.77441406	3818.82568359
bitcoin	2017-08-14	4433.5234375	4356.56689453	4093.12060547
bitcoin	2017-08-15	4537.18408203	4512.39160156	4327.17578125
bitcoin	2017-08-16	4497.26123047	4618.01025391	4482.53271484
bitcoin	2017-08-17	4590.20751953	4577.33203125	4587.84423828
bitcoin	2017-08-18	4460.5546875	4672.04199219	4547.28320312
bitcoin	2017-08-19	4280.82714844	4539.93164062	4641.72021484
bitcoin	2017-08-20	4273.15527344	4356.84130859	4509.9921875
bitcoin	2017-08-21	4184.88037109	4349.02685547	4327.45019531
bitcoin	2017-08-22	4232.03515625	4259.11962891	4319.65917969
bitcoin	2017-08-23	4349.29736328	4307.14453125	4230.02978516
bitcoin	2017-08-24	4461.98828125	4426.58740234	4277.90576172
bitcoin	2017-08-25	4563.40380859	4541.39257812	4396.984375
bitcoin	2017-08-26	4477.74902344	4644.72753906	4511.44873047
bitcoin	2017-08-27	4507.77734375	4557.45117188	4614.48486328
bitcoin	2017-08-28	4502.52978516	4588.04589844	4527.45996094
bitcoin	2017-08-29	4756.88964844	4582.69970703	4557.96630859
bitcoin	2017-08-30	4752.96923828	4841.91992188	4552.63525391
}{\bitcoinPriceAndRLanguage}

\pgfplotstableread{
date	yTrue	coin
2017-05-04	97.66	ethereum
2017-05-05	95.6	ethereum
2017-05-06	96.37	ethereum
2017-05-07	93.45	ethereum
2017-05-08	89.58	ethereum
2017-05-09	91.32	ethereum
2017-05-10	90.55	ethereum
2017-05-11	90.57	ethereum
2017-05-12	88.59	ethereum
2017-05-13	89.7	ethereum
2017-05-14	94.3	ethereum
2017-05-15	92.18	ethereum
2017-05-16	89.4	ethereum
2017-05-17	96.1	ethereum
2017-05-18	129.47	ethereum
2017-05-19	129.07	ethereum
2017-05-20	148.6	ethereum
2017-05-21	181.46	ethereum
2017-05-22	176.04	ethereum
2017-05-23	201.15	ethereum
2017-05-24	216.1	ethereum
2017-05-25	197.67	ethereum
2017-05-26	169.62	ethereum
2017-05-27	182.36	ethereum
2017-05-28	197.74	ethereum
2017-05-29	232.08	ethereum
2017-05-30	233.92	ethereum
2017-05-31	235.52	ethereum
2017-06-01	227.71	ethereum
2017-06-02	226.41	ethereum
2017-06-03	249.34	ethereum
2017-06-04	248.62	ethereum
2017-06-05	268.12	ethereum
2017-06-06	264.8	ethereum
2017-06-07	260.77	ethereum
2017-06-08	279.14	ethereum
2017-06-09	351.82	ethereum
2017-06-10	351.21	ethereum
2017-06-11	417.21	ethereum
2017-06-12	399.5	ethereum
2017-06-13	393.62	ethereum
2017-06-14	351.94	ethereum
2017-06-15	360.84	ethereum
2017-06-16	368.33	ethereum
2017-06-17	379.17	ethereum
2017-06-18	361.5	ethereum
2017-06-19	365.22	ethereum
2017-06-20	352.83	ethereum
2017-06-21	333.4	ethereum
2017-06-22	331.53	ethereum
2017-06-23	330.28	ethereum
2017-06-24	309.98	ethereum
2017-06-25	293.55	ethereum
2017-06-26	286.64	ethereum
2017-06-27	323.87	ethereum
2017-06-28	322.28	ethereum
2017-06-29	305.3	ethereum
2017-06-30	281.81	ethereum
2017-07-01	293.3	ethereum
2017-07-02	284.31	ethereum
2017-07-03	282.32	ethereum
2017-07-04	274.03	ethereum
2017-07-05	274.55	ethereum
2017-07-06	266.25	ethereum
2017-07-07	248.35	ethereum
2017-07-08	251.79	ethereum
2017-07-09	240.09	ethereum
2017-07-10	216.16	ethereum
2017-07-11	227.06	ethereum
2017-07-12	226.66	ethereum
2017-07-13	207.07	ethereum
2017-07-14	198.71	ethereum
2017-07-15	171.91	ethereum
2017-07-16	190.26	ethereum
2017-07-17	256.89	ethereum
2017-07-18	244.02	ethereum
2017-07-19	235.86	ethereum
2017-07-20	236.62	ethereum
2017-07-21	236.55	ethereum
2017-07-22	234.17	ethereum
2017-07-23	229.56	ethereum
2017-07-24	225.98	ethereum
2017-07-25	209.29	ethereum
2017-07-26	205.68	ethereum
2017-07-27	203.93	ethereum
2017-07-28	209.57	ethereum
2017-07-29	209.88	ethereum
2017-07-30	201.78	ethereum
2017-07-31	232.55	ethereum
2017-08-01	228.98	ethereum
2017-08-02	228.08	ethereum
2017-08-03	227.45	ethereum
2017-08-04	258.47	ethereum
2017-08-05	272.22	ethereum
2017-08-06	274.39	ethereum
2017-08-07	299.55	ethereum
2017-08-08	314.75	ethereum
2017-08-09	310.18	ethereum
2017-08-10	310.36	ethereum
2017-08-11	319.14	ethereum
2017-08-12	308.87	ethereum
2017-08-13	306.3	ethereum
2017-08-14	299.71	ethereum
2017-08-15	303.29	ethereum
2017-08-16	310.43	ethereum
2017-08-17	306.52	ethereum
2017-08-18	298.79	ethereum
2017-08-19	298.78	ethereum
2017-08-20	345.44	ethereum
2017-08-21	329.39	ethereum
2017-08-22	325.1	ethereum
2017-08-23	328.65	ethereum
2017-08-24	337.24	ethereum
2017-08-25	334.7	ethereum
2017-08-26	347.92	ethereum
2017-08-27	350.5	ethereum
2017-08-28	375.18	ethereum
2017-08-29	390.64	ethereum
2017-08-30	389.64	ethereum

}{\ethereumTrue}

\pgfplotstableread{
coin	date	oneDay	twoDay	threeDay
ethereum	2017-05-04	94.97805786129999	80.8381195068	81.2415313721
ethereum	2017-05-05	97.30331420899999	95.0047683716	80.9637451172
ethereum	2017-05-06	95.2650146484	97.3240661621	95.1960754395
ethereum	2017-05-07	96.0269165039	95.29097747799999	97.52596282959999
ethereum	2017-05-08	93.13753509520001	96.0509338379	95.4836044312
ethereum	2017-05-09	89.3077468872	93.1689147949	96.2470397949
ethereum	2017-05-10	91.0297164917	89.34879302979999	93.35182189940001
ethereum	2017-05-11	90.26770019530001	91.0664367676	89.5140991211
ethereum	2017-05-12	90.2874984741	90.30634307860001	91.23965454100001
ethereum	2017-05-13	88.3279800415	90.3260803223	90.4760513306
ethereum	2017-05-14	89.4265060425	88.37147521969999	90.4959030151
ethereum	2017-05-15	93.9786453247	89.46725463870001	88.5322494507
ethereum	2017-05-16	91.8807830811	94.0078887939	89.633102417
ethereum	2017-05-17	89.1296157837	91.91533660889999	94.19464874270001
ethereum	2017-05-18	95.7597579956	89.1710891724	92.0924911499
ethereum	2017-05-19	128.760543823	95.7844696045	89.3355789185
ethereum	2017-05-20	128.365219116	128.696182251	95.97933197020001
ethereum	2017-05-21	147.658981323	128.301971436	129.036300659
ethereum	2017-05-22	180.07791137700002	147.53918457	128.640441895
ethereum	2017-05-23	174.73486328099997	179.854736328	147.957702637
ethereum	2017-05-24	199.472518921	174.52949523900003	180.399047852
ethereum	2017-05-25	214.180404663	199.182174683	175.053848267
ethereum	2017-05-26	196.046585083	213.83627319299998	199.796188354
ethereum	2017-05-27	168.403823853	195.768417358	214.500244141
ethereum	2017-05-28	180.964935303	168.21910095200002	196.370422363
ethereum	2017-05-29	196.115509033	180.73878479	168.719421387
ethereum	2017-05-30	229.88301086400003	195.83709716799999	181.286376953
ethereum	2017-05-31	231.68978881799998	229.478683472	196.439346313
ethereum	2017-06-01	233.260650635	231.278320312	230.19314575200002
ethereum	2017-06-02	225.59085083	232.84298706099997	231.99836731
ethereum	2017-06-03	224.313735962	225.20333862299998	233.567886353
ethereum	2017-06-04	246.82005310099998	223.93112182599998	225.90423584
ethereum	2017-06-05	246.11404418900003	246.34744262700002	224.62799072299998
ethereum	2017-06-06	265.218841553	245.64433288599997	247.112838745
ethereum	2017-06-07	261.96862793	264.667938232	246.407730103
ethereum	2017-06-08	258.02194213900003	261.431854248	265.484588623
ethereum	2017-06-09	275.999755859	257.502166748	262.239746094
ethereum	2017-06-10	346.7784729	275.400878906	258.29928588900003
ethereum	2017-06-11	346.186981201	345.823944092	276.245452881
ethereum	2017-06-12	409.89514160199997	345.235778809	346.81158447300004
ethereum	2017-06-13	392.85949707	408.55798339800003	346.222473145
ethereum	2017-06-14	387.193450928	391.63214111300005	409.608886719
ethereum	2017-06-15	346.894805908	386.00152587900004	392.672271729
ethereum	2017-06-16	355.519042969	345.93966674800004	387.037017822
ethereum	2017-06-17	362.76907348599997	354.515441895	346.927429199
ethereum	2017-06-18	373.24865722699997	361.7237854	355.515533447
ethereum	2017-06-19	356.158172607	372.14163208	362.73339843800005
ethereum	2017-06-20	359.75958252	355.150909424	373.163513184
ethereum	2017-06-21	347.757720947	358.73171997099996	356.15188598599997
ethereum	2017-06-22	328.897613525	346.79776001	359.737457275
ethereum	2017-06-23	327.08001709	328.03991699200003	347.786834717
ethereum	2017-06-24	325.86486816400003	326.231933594	328.998199463
ethereum	2017-06-25	306.10510253900003	325.023101807	327.187011719
ethereum	2017-06-26	290.078918457	305.363861084	325.975982666
ethereum	2017-06-27	283.330169678	289.415008545	306.27804565400004
ethereum	2017-06-28	319.630554199	282.697784424	290.293701172
ethereum	2017-06-29	318.083404541	318.821136475	283.56048584
ethereum	2017-06-30	301.543029785	317.281890869	319.762451172
ethereum	2017-07-01	278.610046387	300.824188232	318.22024536099997
ethereum	2017-07-02	289.834777832	277.999267578	301.728668213
ethereum	2017-07-03	281.053466797	289.172027588	278.850402832
ethereum	2017-07-04	279.108551025	280.43157959	290.050170898
ethereum	2017-07-05	271.002105713	278.495544434	281.28869628900003
ethereum	2017-07-06	271.510772705	270.425598145	279.347900391
ethereum	2017-07-07	263.388275146	270.93206787099996	271.257476807
ethereum	2017-07-08	245.84925842299998	262.84533691400003	271.76519775400004
ethereum	2017-07-09	249.22213745099998	245.380691528	263.657073975
ethereum	2017-07-10	237.746276855	248.739547729	246.143249512
ethereum	2017-07-11	214.239395142	237.310684204	249.511886597
ethereum	2017-07-12	224.952316284	213.895065308	238.04922485400002
ethereum	2017-07-13	224.55934143099998	224.567214966	214.55923461900002
ethereum	2017-07-14	205.298583984	224.17578125	225.26612854
ethereum	2017-07-15	197.07052612299998	204.987213135	224.87344360400002
ethereum	2017-07-16	170.662368774	196.78872680700002	205.621337891
ethereum	2017-07-17	188.74900817900001	170.470306396	197.39433288599997
ethereum	2017-07-18	254.220672607	188.49632263200002	170.979278564
ethereum	2017-07-19	241.60232543900003	253.71705627400002	189.07231140099998
ethereum	2017-07-20	233.594436646	241.151092529	254.503662109
ethereum	2017-07-21	234.340530396	233.17543029799998	241.90121459999997
ethereum	2017-07-22	234.27178955099998	233.918548584	233.901367188
ethereum	2017-07-23	231.935272217	233.85009765599997	234.64674377400002
ethereum	2017-07-24	227.408111572	231.522842407	234.578094482
ethereum	2017-07-25	223.891281128	227.01347351099997	232.243667603
ethereum	2017-07-26	207.482696533	223.510238647	227.720123291
ethereum	2017-07-27	203.93084716799999	207.163375854	224.205749512
ethereum	2017-07-28	202.20867919900002	203.624465942	207.804931641
ethereum	2017-07-29	207.75814819299998	201.90853881799998	204.253890991
ethereum	2017-07-30	208.063110352	207.43783569299998	202.53204345700001
ethereum	2017-07-31	200.092636108	207.741653442	208.08029174799998
ethereum	2017-08-01	230.344558716	199.800064087	208.385147095
ethereum	2017-08-02	226.838424683	229.938446045	200.41624450700002
ethereum	2017-08-03	225.954345703	226.445999146	230.654312134
ethereum	2017-08-04	225.33544921900003	225.56536865200002	227.15083313
ethereum	2017-08-05	255.76878356900002	224.94886779799998	226.267456055
ethereum	2017-08-06	269.231262207	255.25860595700001	225.649002075
ethereum	2017-08-07	271.354278564	268.662658691	256.04953002900004
ethereum	2017-08-08	295.934753418	270.7762146	269.489929199
ethereum	2017-08-09	310.75238037099996	295.243011475	271.608978271
ethereum	2017-08-10	306.299987793	309.987976074	296.13513183599997
ethereum	2017-08-11	306.47543335	305.557769775	310.911773682
ethereum	2017-08-12	315.027160645	305.732299805	306.47238159200003
ethereum	2017-08-13	305.023284912	314.24118042	306.647247314
ethereum	2017-08-14	302.518096924	304.28738403299997	315.173583984
ethereum	2017-08-15	296.090881348	301.79446411099997	305.199310303
ethereum	2017-08-16	299.583007812	295.398345947	302.70098877
ethereum	2017-08-17	306.543609619	298.873626709	296.290863037
ethereum	2017-08-18	302.732543945	305.800201416	299.77389526400003
ethereum	2017-08-19	295.19317627	302.007843018	306.71530151400003
ethereum	2017-08-20	295.18347168	294.50500488299997	302.914886475
ethereum	2017-08-21	340.589874268	294.495300293	295.395446777
ethereum	2017-08-22	324.99957275400004	339.669433594	295.385742188
ethereum	2017-08-23	320.827178955	324.162322998	340.647460938
ethereum	2017-08-24	324.27999877900004	320.011657715	325.11364746099997
ethereum	2017-08-25	332.628601074	323.446502686	320.95513916
ethereum	2017-08-26	330.160919189	331.75112915	324.39654541
ethereum	2017-08-27	342.99609375	329.29650878900003	332.71594238299997
ethereum	2017-08-28	345.49847412099996	342.0625	330.257080078
ethereum	2017-08-29	369.393127441	344.551116943	343.044250488
ethereum	2017-08-30	384.319976807	368.309082031	345.536712646
}{\ethereumPriceAndRLanguage}

\pgfplotstableread{
date	yTrue	coin
2017-05-04	31.48	monero
2017-05-05	29.92	monero
2017-05-06	37.97	monero
2017-05-07	33.14	monero
2017-05-08	30.79	monero
2017-05-09	30.94	monero
2017-05-10	30.86	monero
2017-05-11	30.32	monero
2017-05-12	28.87	monero
2017-05-13	29.01	monero
2017-05-14	28.62	monero
2017-05-15	27.79	monero
2017-05-16	28.34	monero
2017-05-17	31.3	monero
2017-05-18	32.0	monero
2017-05-19	36.99	monero
2017-05-20	36.51	monero
2017-05-21	40.41	monero
2017-05-22	55.97	monero
2017-05-23	58.5	monero
2017-05-24	51.61	monero
2017-05-25	45.0	monero
2017-05-26	40.29	monero
2017-05-27	39.89	monero
2017-05-28	43.95	monero
2017-05-29	46.21	monero
2017-05-30	43.92	monero
2017-05-31	45.86	monero
2017-06-01	45.0	monero
2017-06-02	44.91	monero
2017-06-03	43.72	monero
2017-06-04	48.68	monero
2017-06-05	54.99	monero
2017-06-06	61.69	monero
2017-06-07	56.62	monero
2017-06-08	57.41	monero
2017-06-09	56.66	monero
2017-06-10	57.72	monero
2017-06-11	59.26	monero
2017-06-12	52.79	monero
2017-06-13	53.83	monero
2017-06-14	46.95	monero
2017-06-15	48.18	monero
2017-06-16	52.61	monero
2017-06-17	54.03	monero
2017-06-18	51.44	monero
2017-06-19	53.53	monero
2017-06-20	52.16	monero
2017-06-21	50.53	monero
2017-06-22	51.96	monero
2017-06-23	51.0	monero
2017-06-24	48.09	monero
2017-06-25	47.3	monero
2017-06-26	44.98	monero
2017-06-27	47.55	monero
2017-06-28	47.75	monero
2017-06-29	45.49	monero
2017-06-30	43.57	monero
2017-07-01	42.72	monero
2017-07-02	44.17	monero
2017-07-03	45.72	monero
2017-07-04	46.66	monero
2017-07-05	50.16	monero
2017-07-06	50.66	monero
2017-07-07	46.02	monero
2017-07-08	46.97	monero
2017-07-09	45.22	monero
2017-07-10	41.37	monero
2017-07-11	40.92	monero
2017-07-12	40.62	monero
2017-07-13	38.42	monero
2017-07-14	35.43	monero
2017-07-15	32.63	monero
2017-07-16	35.0	monero
2017-07-17	37.98	monero
2017-07-18	37.88	monero
2017-07-19	41.63	monero
2017-07-20	42.56	monero
2017-07-21	44.94	monero
2017-07-22	44.96	monero
2017-07-23	46.07	monero
2017-07-24	46.26	monero
2017-07-25	46.93	monero
2017-07-26	46.66	monero
2017-07-27	45.87	monero
2017-07-28	44.39	monero
2017-07-29	44.14	monero
2017-07-30	41.16	monero
2017-07-31	45.24	monero
2017-08-01	48.23	monero
2017-08-02	44.54	monero
2017-08-03	45.61	monero
2017-08-04	50.02	monero
2017-08-05	49.51	monero
2017-08-06	50.23	monero
2017-08-07	52.85	monero
2017-08-08	53.46	monero
2017-08-09	51.38	monero
2017-08-10	51.49	monero
2017-08-11	51.76	monero
2017-08-12	50.14	monero
2017-08-13	49.93	monero
2017-08-14	50.26	monero
2017-08-15	49.05	monero
2017-08-16	49.44	monero
2017-08-17	49.0	monero
2017-08-18	58.31	monero
2017-08-19	56.2	monero
2017-08-20	98.25	monero
2017-08-21	95.77	monero
2017-08-22	100.08	monero
2017-08-23	94.12	monero
2017-08-24	108.8	monero
2017-08-25	148.19	monero
2017-08-26	139.18	monero
2017-08-27	154.88	monero
2017-08-28	145.57	monero
2017-08-29	137.76	monero
2017-08-30	143.67	monero
}{\moneroTrue}

\pgfplotstableread{
coin	date	oneDay	twoDay	threeDay
monero	2017-05-04	27.092571258499998	26.012180328400003	23.7765045166
monero	2017-05-05	31.895488739	27.1529083252	26.192274093600002
monero	2017-05-06	30.301288604699998	32.0267829895	27.3041095734
monero	2017-05-07	38.5279808044	30.4091072083	32.0544624329
monero	2017-05-08	33.5919075012	38.7561569214	30.4778003693
monero	2017-05-09	31.1903591156	33.7481079102	38.6131401062
monero	2017-05-10	31.3436470032	31.3112754822	33.7321395874
monero	2017-05-11	31.2618923187	31.466823577899998	31.357099533099998
monero	2017-05-12	30.7100582123	31.3838615417	31.5086994171
monero	2017-05-13	29.2282924652	30.823894500700003	31.427846908600003
monero	2017-05-14	29.3713569641	29.3202610016	30.882076263400002
monero	2017-05-15	28.972822189299997	29.4654407501	29.41655159
monero	2017-05-16	28.1246585846	29.061010360700003	29.5580558777
monero	2017-05-17	28.6866931915	28.200290679899997	29.1638717651
monero	2017-05-18	31.7115383148	28.7706508636	28.3249664307
monero	2017-05-19	32.4268875122	31.8401260376	28.880868911700002
monero	2017-05-20	37.5264511108	32.565998077399996	31.872539520300002
monero	2017-05-21	37.0359039307	37.7400856018	32.5800056458
monero	2017-05-22	41.0215835571	37.2424049377	37.6228523254
monero	2017-05-23	56.9218406677	41.2858314514	37.137798309299995
monero	2017-05-24	59.506637573199995	57.4111213684	41.0786323547
monero	2017-05-25	52.4669647217	60.0316238403	56.795001983599995
monero	2017-05-26	45.7123374939	52.894115448	59.349174499499995
monero	2017-05-27	40.8989486694	46.043895721400006	52.3924407959
monero	2017-05-28	40.4901580811	41.1614265442	45.7159576416
monero	2017-05-29	44.6393051147	40.746738433800004	40.9573783875
monero	2017-05-30	46.9488677979	44.9555244446	40.5532150269
monero	2017-05-31	44.6086502075	47.298034668	44.6552085876
monero	2017-06-01	46.5911941528	44.9244232178	46.93828964229999
monero	2017-06-02	45.7123374939	46.9352798462	44.6249008179
monero	2017-06-03	45.6203613281	46.043895721400006	46.584732055699995
monero	2017-06-04	44.4042549133	45.9506111145	45.7159576416
monero	2017-06-05	49.4729499817	44.7171134949	45.6250343323
monero	2017-06-06	55.920562744099996	49.8579177856	44.422843933100005
monero	2017-06-07	62.7654190063	56.395942688000005	49.4332389832
monero	2017-06-08	57.5859375	63.335041046099995	55.8055305481
monero	2017-06-09	58.393054962200004	58.084419250500005	62.5690345764
monero	2017-06-10	57.6268081665	58.9026870728	57.451259613000005
monero	2017-06-11	58.709762573199995	58.125846862799996	58.2488059998
monero	2017-06-12	60.2830619812	59.2237701416	57.4916305542
monero	2017-06-13	53.672695159899995	60.818717956499995	58.5617637634
monero	2017-06-14	54.7353363037	54.1167297363	60.116340637200004
monero	2017-06-15	47.705085754399995	55.194213867200006	53.584064483599995
monero	2017-06-16	48.9620094299	48.065002441400004	54.6342468262
monero	2017-06-17	53.4887695312	49.3397483826	47.685798645
monero	2017-06-18	54.9396896362	53.930229186999995	48.928211212200004
monero	2017-06-19	52.293254852299995	55.4014129639	53.4023017883
monero	2017-06-20	54.4288063049	52.717964172399995	54.8362045288
monero	2017-06-21	53.028961181599996	54.8834037781	52.220760345500004
monero	2017-06-22	51.3633804321	53.463989257799994	54.331314086899994
monero	2017-06-23	52.8245925903	51.775020599399994	52.9478683472
monero	2017-06-24	51.8436546326	53.2567596436	51.3017234802
monero	2017-06-25	48.8700332642	52.2620391846	52.7458992004
monero	2017-06-26	48.0627365112	49.2464752197	51.7763977051
monero	2017-06-27	45.6918945312	48.4277458191	48.8373069763
monero	2017-06-28	48.3182144165	46.0231628418	48.039333343500005
monero	2017-06-29	48.5225944519	48.686836242700004	45.6957588196
monero	2017-06-30	46.2130851746	48.894123077399996	48.2918586731
monero	2017-07-01	44.2509651184	46.551776886000006	48.493881225600006
monero	2017-07-02	43.382312774700004	44.561630249	46.210956573500006
monero	2017-07-03	44.864135742200006	43.6805229187	44.2713088989
monero	2017-07-04	46.448120117200006	45.1835708618	43.4125556946
monero	2017-07-05	47.4087257385	46.790168762200004	44.877460479700005
monero	2017-07-06	50.9853057861	47.7644348145	46.443302154499996
monero	2017-07-07	51.496223449700004	51.391620636000006	47.3928565979
monero	2017-07-08	46.754699707	51.9097366333	50.928047180200004
monero	2017-07-09	47.7255134583	47.101108551	51.433017730699994
monero	2017-07-10	45.937164306599996	48.0857276917	46.7463607788
monero	2017-07-11	42.0026702881	46.2719268799	47.705993652299995
monero	2017-07-12	41.5427856445	42.2810554504	45.938205718999995
monero	2017-07-13	41.2361984253	41.8145484924	42.0485954285
monero	2017-07-14	38.987865448	41.503540039099995	41.5939331055
monero	2017-07-15	35.932182311999995	39.222709655799996	41.290813446
monero	2017-07-16	33.0707130432	36.1226005554	39.0678596497
monero	2017-07-17	35.4927406311	33.2192687988	36.0464057922
monero	2017-07-18	38.5382003784	35.6767425537	33.2167129517
monero	2017-07-19	38.4360046387	38.7665214539	35.6118583679
monero	2017-07-20	42.2683753967	38.6628456116	38.623249054
monero	2017-07-21	43.2187957764	42.5505867004	38.5222015381
monero	2017-07-22	45.6510238647	43.5146636963	42.3112945557
monero	2017-07-23	45.6714630127	45.981704711899994	43.2509078979
monero	2017-07-24	46.8057899475	46.002429962200004	45.6553497314
monero	2017-07-25	46.9999656677	47.1529312134	45.675552368199995
monero	2017-07-26	47.6846389771	47.349861145	46.7968711853
monero	2017-07-27	47.4087257385	48.044273376499994	46.9888000488
monero	2017-07-28	46.6014099121	47.7644348145	47.6655960083
monero	2017-07-29	45.088958740200006	46.945644378699996	47.3928565979
monero	2017-07-30	44.8334732056	45.411609649700004	46.594833374
monero	2017-07-31	41.7880592346	45.152473449700004	45.0997161865
monero	2017-08-01	45.957599639899996	42.0633506775	44.8471641541
monero	2017-08-02	49.013103485100004	46.292655944799996	41.8364219666
monero	2017-08-03	45.2422523499	49.3915710449	45.9584197998
monero	2017-08-04	46.3357124329	45.5670890808	48.9787101746
monero	2017-08-05	50.842250824	46.6761627197	45.251247406000005
monero	2017-08-06	50.321098327600005	51.246547699	46.3321838379
monero	2017-08-07	51.056831359899995	50.7180519104	50.7866477966
monero	2017-08-08	53.7339935303	51.464164733900006	50.2715530396
monero	2017-08-09	54.3572807312	54.1788902283	50.9987449646
monero	2017-08-10	52.231945037799996	54.810878753699996	53.644653320299994
monero	2017-08-11	52.3443450928	52.6557922363	54.2606391907
monero	2017-08-12	52.6202316284	52.7697639465	52.1601600647
monero	2017-08-13	50.9648666382	53.0495300293	52.2712516785
monero	2017-08-14	50.7502746582	51.370895385699995	52.543926238999994
monero	2017-08-15	51.08749389649999	51.1532821655	50.907844543500005
monero	2017-08-16	49.8510398865	51.4952430725	50.695751190200006
monero	2017-08-17	50.249561309799994	50.241355896	51.029045105
monero	2017-08-18	49.7999458313	50.645515441899995	49.8069419861
monero	2017-08-19	59.3125305176	50.189544677700006	50.200859069799996
monero	2017-08-20	57.1568336487	59.83484268189999	49.7564544678
monero	2017-08-21	100.070655823	57.6493721008	59.1573638916
monero	2017-08-22	97.54349517819999	101.11743164100001	57.027214050299996
monero	2017-08-23	101.935050964	98.5601196289	99.4064483643
monero	2017-08-24	95.8618087769	103.003852844	96.9122390747
monero	2017-08-25	110.81391143799999	96.85816192629999	101.246421814
monero	2017-08-26	150.791290283	111.984924316	95.25235748290001
monero	2017-08-27	141.668731689	152.361923218	110.007675171
monero	2017-08-28	157.555206299	143.157562256	149.42991638200002
monero	2017-08-29	148.140090942	159.182571411	140.437561035
monero	2017-08-30	140.229690552	149.687530518	156.095947266
}{\moneroPriceAndRLanguage}

\begin{tikzpicture}
\begin{axis}[
title = {\small $\$_{High}$ in 1 day},title style ={yshift=-0.1in},
ylabel = Bitcoin, 
height = 1.25 in, width = 2.5 in,
height = 1.3in,
xtick align = center,date coordinates in=x, xtick={2017-05-04,2017-06-01,2017-07-01,2017-08-01,2017-09-01},
xticklabels={, , , , }, 
ytick align = center,
xticklabel style={align=left},
ytick ={2000,3000,4000,5000}, yticklabels={2K~~, 3K~~, 4K~~, 5K~~},
xlabel style = {yshift = -0.2 in}, xtick pos= left,
legend style={draw=none,fill=none,font=\tiny,legend cell align=left,legend columns = 1, column sep = 0.2mm}]

\addplot[black!50!white, no marks]
table [x=date, y=yTrue] {\bitcoinTrue};

\addplot[reddit,  mark= o, mark size=0.5] table [x=date, y= oneDay] {\bitcoinPriceAndRLanguage};  

\end{axis}
\end{tikzpicture}
\begin{tikzpicture}
\begin{axis}[
title = {\small $\$_{High}$ in 2 days},title style ={yshift=-0.1in}, 
height = 1.25 in, width = 2.5 in,
height = 1.3in,
xtick align = center,date coordinates in=x, xtick={2017-05-04,2017-06-01,2017-07-01,2017-08-01,2017-09-01},
xticklabels={, , , , },  
ytick align = center,
xticklabel style={align=left},
ytick ={2000,3000,4000,5000}, yticklabels={, , , },
xlabel style = {yshift = -0.2 in}, xtick pos= left, 
]

\addplot[black!50!white, no marks]
table [x=date, y=yTrue] {\bitcoinTrue};

\addplot[reddit,  mark= o, mark size=0.5] table [x=date, y= twoDay] {\bitcoinPriceAndRLanguage};  

\end{axis}
\end{tikzpicture}
\begin{tikzpicture}
\begin{axis}[
title = {\small $\$_{High}$ in 3 days},title style ={yshift=-0.1in}, 
height = 1.25 in, width = 2.5 in,
height = 1.3in,
xtick align = center,date coordinates in=x, xtick={2017-05-04,2017-06-01,2017-07-01,2017-08-01,2017-09-01},
xticklabels={, , , , },  
ytick align = center,
xticklabel style={align=left},
ytick ={2000,3000,4000,5000}, yticklabels={, , , },
xlabel style = {yshift = -0.2 in}, xtick pos= left, 
]

\addplot[black!50!white, no marks]
table [x=date, y=yTrue] {\bitcoinTrue};

\addplot[reddit,  mark= o, mark size=0.5] table [x=date, y= threeDay] {\bitcoinPriceAndRLanguage};  

\end{axis}
\end{tikzpicture}

\begin{tikzpicture}
\begin{axis}[ 
height = 1.25 in, width = 2.5 in,
height = 1.3in,
ylabel = Ethereum, 
xtick align = center,date coordinates in=x, xtick={2017-05-04,2017-06-01,2017-07-01,2017-08-01,2017-09-01},
xticklabels={, , , , },  
ytick align = center,
xticklabel style={align=left},
ytick ={100,200,300,400}, yticklabels={100,200,300,400},
xlabel style = {yshift = -0.2 in}, xtick pos= left, 
]

\addplot[black!50!white, no marks] 
table [x=date, y=yTrue] {\ethereumTrue};

\addplot[reddit,  mark= o, mark size=0.5] table [x=date, y= oneDay] {\ethereumPriceAndRLanguage};  

\end{axis}
\end{tikzpicture}
\begin{tikzpicture}
\begin{axis}[ 
height = 1.25 in, width = 2.5 in,
height = 1.3in,
xtick align = center,date coordinates in=x, xtick={2017-05-04,2017-06-01,2017-07-01,2017-08-01,2017-09-01},
xticklabels={, , , , },  
ytick align = center,
xticklabel style={align=left},
ytick ={100,200,300,400}, yticklabels={, , , },
xlabel style = {yshift = -0.2 in}, xtick pos= left, 
]

\addplot[black!50!white, no marks] 
table [x=date, y=yTrue] {\ethereumTrue};

\addplot[reddit,  mark= o, mark size=0.5] table [x=date, y= twoDay] {\ethereumPriceAndRLanguage};  

\end{axis}
\end{tikzpicture}
\begin{tikzpicture}
\begin{axis}[ 
height = 1.25 in, width = 2.5 in,
height = 1.3in,
xtick align = center,date coordinates in=x, xtick={2017-05-04,2017-06-01,2017-07-01,2017-08-01,2017-09-01},
xticklabels={, , , , },  
ytick align = center,
xticklabel style={align=left},
ytick ={100,200,300,400}, yticklabels={, , , },
xlabel style = {yshift = -0.2 in}, xtick pos= left, 
]

\addplot[black!50!white, no marks] 
table [x=date, y=yTrue] {\ethereumTrue};

\addplot[reddit,  mark= o, mark size=0.5] table [x=date, y= threeDay] {\ethereumPriceAndRLanguage};  

\end{axis}
\end{tikzpicture}

\begin{tikzpicture}
\begin{axis}[ 
height = 1.25 in, width = 2.5 in,
height = 1.3in,
ylabel = Monero, 
xtick align = center,date coordinates in=x, xtick={2017-05-04,2017-06-01,2017-07-01,2017-08-01,2017-09-01},
xticklabels={, , , , },  
ytick align = center,
xticklabel style={align=left}, 
xlabel style = {yshift = -0.2 in}, xtick pos= left, 
extra x ticks={2017-05-18,2017-06-15,2017-07-16,2017-08-16},
extra x tick labels={May\\2017, June\\2017, July\\2017, August\\2017}, 
every extra x tick/.style={tickwidth=0,xticklabel style={align=center}},
ytick={50,100,150}, yticklabels={50,100,150}, 
xlabel style = {yshift = -0.2 in}, xtick pos= left,
]

\addplot[black!50!white, no marks] 
table [x=date, y=yTrue] {\moneroTrue};

\addplot[reddit,  mark= o, mark size=0.5] table [x=date, y= oneDay] {\moneroPriceAndRLanguage};  

\end{axis}
\end{tikzpicture}
\begin{tikzpicture}
\begin{axis}[ 
height = 1.25 in, width = 2.5 in,
height = 1.3in,
xtick align = center,date coordinates in=x, xtick={2017-05-04,2017-06-01,2017-07-01,2017-08-01,2017-09-01},
xticklabels={, , , , },  
ytick align = center,
xticklabel style={align=left},
ytick ={100,200,300,400}, yticklabels={, , , },
xlabel style = {yshift = -0.2 in}, xtick pos= left, 
extra x ticks={2017-05-18,2017-06-15,2017-07-16,2017-08-16},
extra x tick labels={May\\2017, June\\2017, July\\2017, August\\2017}, 
every extra x tick/.style={tickwidth=0,xticklabel style={align=center}},
ytick={50,100,150}, yticklabels={,,},
xlabel style = {yshift = -0.2 in}, xtick pos= left,
]

\addplot[black!50!white, no marks] 
table [x=date, y=yTrue] {\moneroTrue};

\addplot[reddit,  mark= o, mark size=0.5] table [x=date, y= twoDay] {\moneroPriceAndRLanguage};  

\end{axis}
\end{tikzpicture}
\begin{tikzpicture}
\begin{axis}[ 
height = 1.25 in, width = 2.5 in,
height = 1.3in,
xtick align = center,date coordinates in=x, xtick={2017-05-04,2017-06-01,2017-07-01,2017-08-01,2017-09-01},
xticklabels={, , , , },  
ytick align = center,
xticklabel style={align=left},
ytick ={50,100,150}, yticklabels={, , },
xlabel style = {yshift = -0.2 in}, xtick pos= left, 
extra x ticks={2017-05-18,2017-06-15,2017-07-16,2017-08-16},
extra x tick labels={May\\2017, June\\2017, July\\2017, August\\2017}, 
every extra x tick/.style={tickwidth=0,xticklabel style={align=center}},
ytick={50,100,150}, yticklabels={,,},
xlabel style = {yshift = -0.2 in}, xtick pos= left,
]

\addplot[black!50!white, no marks] 
table [x=date, y=yTrue] {\moneroTrue};

\addplot[reddit,  mark= o, mark size=0.5] table [x=date, y= threeDay] {\moneroPriceAndRLanguage};  

\end{axis}
\end{tikzpicture}

\begin{tikzpicture} 
\begin{axis}[%
hide axis,
xmin=10,xmax=50,ymin=0,ymax=0.4,
legend style={draw=none,legend cell align=left,legend columns = -1, column sep = 3mm}
]  
\addlegendimage{black!50!white, no marks,solid,thick}; 
\addlegendentry{True Price (USD)};  
\addlegendimage{reddit, mark=*, mark size=0.75, very thick};  
\addlegendentry{Price (USD) Predictions using LSTM ($\$+R_{Lang}$)};  
\end{axis}
\end{tikzpicture}  

%% file: figs/best_PplusSM_models_predictions/mape_by_genesis_block_date.tex
 \small
 \hspace{-0.15in}
 \begin{tikzpicture}
 \begin{axis}[
 title = {\small $\$_{High}$ in 1 day},title style = {yshift = -0.1 in}, 
 xlabel= \textcolor{white}{Lifetime  (Days)},
 ylabel =  MaxAPE , ylabel style = {align=center},
 height = 1.2 in, width = 1.5 in,
 xtick align = center,ytick align = center,
 xtick pos= left,ytick pos= left,
 xmin = 0, xmax = 3500,    
 ymin = 15, ymax = 55,  ytick={15,25,35,45,55},
 xtick={0,1000,2000,3000},xticklabels={~~0,1K,2K,3K},
 ]

 \addplot[arima, only marks, mark = +, mark size=2] 
 coordinates {
 	(3163,23.19) 
 	(764,48.19) 
 	(1232,42.71) 
 };
 
 \addplot[price, only marks, mark = x, mark size=2] 
 coordinates {
 	(3163,17.69) 
 	(764,26.8) 
 	(1232,43.01) 
 };
 \addplot[reddit, only marks, mark = *o, mark size=2] 
 coordinates {
 	(1232, 42.81) 
 };
 \addplot[githubAndReddit, only marks, mark = o, mark size=2] 
 coordinates {
 	(3163, 15.74) 
 	(764, 24.81) 
 };
 \end{axis}
 \end{tikzpicture}
 %
 \hspace{-0.08in}
 \begin{tikzpicture}
 \begin{axis}[
 title = {\small $\$_{High}$ in 2 day},title style = {yshift = -0.1 in}, 
 xlabel=Lifetime  (Days),
 height = 1.2 in, width = 1.5 in,
 xtick align = center,ytick align = center,
 xtick pos= left,ytick pos= left,
 xmin = 0, xmax = 3500,    
 ymin = 15, ymax = 55,  ytick={15,25,35,45,55},yticklabels={ , , , , },
 xtick={0,1000,2000,3000},xticklabels={~~0,1K,2K,3K},
 ]

 \addplot[arima, only marks, mark = +, mark size=2] 
 coordinates {
 	(3163,23.35) 
 	(764,48.35) 
 	(1232,41.13) 
 };
 
 \addplot[price, only marks, mark = x, mark size=2] 
 coordinates {
 	(3163,19.62) 
 	(764,33.8) 
 	(1232,42.15) 
 };
 \addplot[github, only marks, mark = o, mark size=2] 
 coordinates {
 	(764, 32.94) 
 	(1232, 40.99) 
 };
 \addplot[githubAndReddit, only marks, mark = o, mark size=2] 
 coordinates {
 	(3163, 17.06) 
 };
 \end{axis}
 \end{tikzpicture}
 %
 \hspace{-0.08in}
 \begin{tikzpicture}
 \begin{axis}[
 title = {\small $\$_{High}$ in 3 days},title style = {yshift = -0.1 in}, 
 xlabel= \textcolor{white}{Lifetime  (Days)},
 height = 1.2 in, width = 1.5 in,
 xtick align = center,ytick align = center,
 xtick pos= left,ytick pos= left,
 xmin = 0, xmax = 3500,    
 ymin = 15, ymax = 55,  ytick={15,25,35,45,55},yticklabels={ , , , , },
 xtick={0,1000,2000,3000},xticklabels={~~0,1K,2K,3K},
 ]

 \addplot[arima, only marks, mark = +, mark size=2] 
 coordinates {
 	(3163,23.94) 
 	(764,49.58) 
 	(1232,49.89) 
 };
 
 \addplot[price, only marks, mark = x, mark size=2] 
 coordinates {
 	(3163,23.55) 
 	(764,37.89) 
 	(1232,50.01) 
 };
 \addplot[github, only marks, mark = o, mark size=2] 
 coordinates {
 };
 \addplot[reddit, only marks, mark = o, mark size=2] 
 coordinates {
 	(764, 36.81) 
 	(1232, 46.70) 
 };
 \addplot[githubAndReddit, only marks, mark = o, mark size=2] 
 coordinates {
 	(3163, 22.66) 
 };

\draw [<-,black!75!white] (3100,26) -- (2900,31);
\draw (2900,30) node [above,black!75!white] {\tiny \textbf{BTC}};
\draw [<-,black!75!white] (764, 30) -- (764,25);
\draw (764,25) node [below,black!75!white] {\tiny \textbf{ETH}};
\draw [<-,black!75!white] (1400,48) -- (2000,48);
\draw (2000,48)  node [right,black!75!white] {\tiny \textbf{XMR}};

 \end{axis}
 
 \end{tikzpicture}

%% file: figs/best_PplusSM_models_predictions/mape_by_marketcap.tex
\hspace{-0.05in}
\hspace{-0.01in}
\begin{tikzpicture}
\begin{axis}[
	xlabel= \textcolor{white}{Market Cap (\$B)},
	ylabel =  MaxAPE , ylabel style = {align=center},
	height = 1.2 in, width = 1.5 in,
	xtick align = center,ytick align = center,
	xtick pos= left,ytick pos= left,
	xmin = -10, xmax = 160, 
	ymin = 15, ymax = 55,  ytick={15,25,35,45,55},
	xtick={0,50,100,150},
	]

\addplot[arima, only marks, mark = +, mark size=2] 
coordinates {
 	(127,23.19) 
 	(29,48.19) 
 	(2,42.71) 
};
 
\addplot[price, only marks, mark = x, mark size=2] 
 coordinates {
	(127,17.69) 
	(29,26.8) 
	(2,43.01) 
};
\addplot[github, only marks, mark = o, mark size=2] 
coordinates {
};
\addplot[reddit, only marks, mark = o, mark size=2] 
coordinates {
	(2, 42.81) 
};
\addplot[githubAndReddit, only marks, mark = o, mark size=2] 
 coordinates {
 	(127, 15.74) 
 	(29, 24.81) 
};
\end{axis}
\end{tikzpicture}
\hspace{-0.18in}
\hspace{0.01in}
\begin{tikzpicture}
\begin{axis}[
xlabel= Market Cap (\$B),
height = 1.2 in, width = 1.5 in,
xtick align = center,ytick align = center,
xtick pos= left,ytick pos= left,
xmin = -10, xmax = 160, 
ymin = 15, ymax = 55,  ytick={15,25,35,45,55},yticklabels={ , , , , },
xtick={0,50,100,150},
]

\addplot[arima, only marks, mark = +, mark size=2] 
coordinates {
	(127,23.35) 
	(29,48.35) 
	(2,41.13) 
};

\addplot[price, only marks, mark = x, mark size=2] 
coordinates {
	(127,19.62) 
	(29,33.8) 
	(2,42.15) 
};
\addplot[github, only marks, mark = o, mark size=2] 
coordinates {
	(29, 32.94) 
	(2, 40.99) 
};
\addplot[reddit, only marks, mark = o, mark size=2] 
coordinates {
};
\addplot[githubAndReddit, only marks, mark = o, mark size=2] 
coordinates {
	(127, 17.06) 
};
\end{axis}
\end{tikzpicture}
\hspace{-0.18in}
\hspace{0.01in}
\begin{tikzpicture}
\begin{axis}[
xlabel= \textcolor{white}{Market Cap (\$B)},
height = 1.2 in, width = 1.5 in,
xtick align = center,ytick align = center,
xtick pos= left,ytick pos= left,
xmin = -10, xmax = 160, 
ymin = 15, ymax = 55,  ytick={15,25,35,45,55},yticklabels={ , , , , },
xtick={0,50,100,150},
]

\addplot[arima, only marks, mark = +, mark size=2] 
coordinates {
	(127,23.94) 
	(29,49.58) 
	(2,49.89) 
};

\addplot[price, only marks, mark = x, mark size=2] 
coordinates {
	(127,23.55) 
	(29,37.89) 
	(2,50.01) 
};
\addplot[github, only marks, mark = o, mark size=2] 
coordinates {
};
\addplot[reddit, only marks, mark = o, mark size=2] 
coordinates {
	(29, 36.81) 
	(2, 46.70) 
};
\addplot[githubAndReddit, only marks, mark = o, mark size=2] 
coordinates {
	(127, 22.66) 
};

\draw [<-,black!75!white] (127,28) -- (127,45);
\draw (127,45) node [above,black!75!white] {\tiny \textbf{BTC}};
\draw [<-,black!75!white] (45,48) -- (65,42);
\draw [<-,black!75!white] (45,38) -- (65,42);
\draw (60,42) node [right,black!75!white] {\tiny \textbf{ETH}};
\draw [<-,black!75!white] (2,42) -- (2,25);
\draw (-10,20)  node [right,black!75!white] {\tiny \textbf{XMR}};

\end{axis}
\end{tikzpicture}

%% file: figs/best_PplusSM_models_predictions/mape_by_std.tex
\hspace{-0.05in}
\hspace{-0.01in}
\begin{tikzpicture}
\begin{axis}[
	ylabel =  MaxAPE , ylabel style = {align=center},
	height = 1.2 in, width = 1.5 in,
	xtick align = center,ytick align = center,
	xtick pos= left,ytick pos= left,
	xmode=log,xmin = 5, xmax = 1500, xtick ={10,100,1000}, 
	ymin = 15, ymax = 55,  ytick={15,25,35,45,55},
	]

\addplot[arima, only marks, mark = +, mark size=2] 
coordinates {
 	(868,23.19) 
 	(99.44,48.19) 
 	(19.71,42.71) 
};
 
\addplot[price, only marks, mark = x, mark size=2] 
 coordinates {
	(868,17.69) 
	(99.44,26.8) 
	(19.71,43.01) 
};
\addplot[github, only marks, mark = o, mark size=2] 
coordinates {
};
\addplot[reddit, only marks, mark = o, mark size=2] 
coordinates {
	(19.71, 42.81) 
};
\addplot[githubAndReddit, only marks, mark = o, mark size=2] 
 coordinates {
 	(868, 15.74) 
 	(99.44, 24.81) 
};
\end{axis}
\end{tikzpicture}
%
\hspace{-0.18in} 
\begin{tikzpicture}
\begin{axis}[
height = 1.2 in, width = 1.5 in,
xtick align = center,ytick align = center,
xtick pos= left,ytick pos= left,
xmode=log,xmin = 5, xmax = 1500, xtick ={10,100,1000}, 
ymin = 15, ymax = 55,  ytick={15,25,35,45,55},yticklabels={ , , , , }, 
]

\addplot[arima, only marks, mark = +, mark size=2] 
coordinates {
	(868,23.35) 
	(99.44,48.35) 
	(19.71,41.13) 
};

\addplot[price, only marks, mark = x, mark size=2] 
coordinates {
	(868,19.62) 
	(99.44,33.8) 
	(19.71,42.15) 
};
\addplot[github, only marks, mark = o, mark size=2] 
coordinates {
	(99.44, 32.94) 
	(19.71, 40.99) 
};
\addplot[reddit, only marks, mark = o, mark size=2] 
coordinates {
};
\addplot[githubAndReddit, only marks, mark = o, mark size=2] 
coordinates {
	(868, 17.06) 
};
\end{axis}
\end{tikzpicture}
%
\hspace{-0.18in}  
\begin{tikzpicture}
\begin{axis}[
height = 1.2 in, width = 1.5 in,
xtick align = center,ytick align = center,
xtick pos= left,ytick pos= left,
xmode=log,xmin = 5, xmax = 1500, xtick ={10,100,1000}, 
ymin = 15, ymax = 55,  ytick={15,25,35,45,55},yticklabels={ , , , , }, 
]

\addplot[arima, only marks, mark = +, mark size=2] 
coordinates {
	(868,23.94) 
	(99.44,49.58) 
	(19.71,49.89) 
};

\addplot[price, only marks, mark = x, mark size=2] 
coordinates {
	(868,23.55) 
	(99.44,37.89) 
	(19.71,50.01) 
};
\addplot[github, only marks, mark = o, mark size=2] 
coordinates {
};
\addplot[reddit, only marks, mark = o, mark size=2] 
coordinates {
	(99.44, 36.81) 
	(19.71, 46.70) 
};
\addplot[githubAndReddit, only marks, mark = o, mark size=2] 
coordinates {
	(868, 22.66) 
};

\draw [<-,black!75!white] (600,23) -- (350,23);
\draw (400,23) node [left,black!75!white] {\tiny \textbf{BTC}};
\draw [<-,black!75!white] (150,48) -- (300,42);
\draw [<-,black!75!white] (150,38) -- (300,42);
\draw (300,42) node [right,black!75!white] {\tiny \textbf{ETH}};
\draw [<-,black!75!white] (19.71,42) -- (19.71,30);
\draw (10,25)  node [right,black!75!white] {\tiny \textbf{XMR}};

\end{axis}
\end{tikzpicture}

~~~~~~~~~~~~Standard Deviation ($\sigma$) of \$ (log)